\documentclass[sigconf]{acmart}

\newif\ifextras
\newif\ifcopyright
\newif\ifack
\newif\ifappendix
\newif\ifdiff
\newif\ifmove

\copyrightfalse
\acktrue
\appendixtrue
\difffalse
\movefalse
\extrastrue

\ifcopyright
\copyrightyear{2020}
\acmYear{2020}
\setcopyright{acmlicensed}\acmConference[CCS '20]{Proceedings of the 2020 ACM SIGSAC Conference on Computer and Communications Security}{November 9--13, 2020}{Virtual Event, USA}
\acmBooktitle{Proceedings of the 2020 ACM SIGSAC Conference on Computer and Communications Security (CCS '20), November 9--13, 2020, Virtual Event, USA}
\acmPrice{15.00}
\acmDOI{10.1145/3372297.3417282}
\acmISBN{978-1-4503-7089-9/20/11}
\else
\setcopyright{none}
\renewcommand\footnotetextcopyrightpermission[1]{}
\fi

\ifdiff
\newcommand{\hl}[1]{{\color{blue} #1}}
\else
\newcommand{\hl}[1]{{\color{black} #1}}
\fi

\ifmove
\newcommand{\move}[1]{{\color{red} #1}}
\else
\newcommand{\move}[1]{}%
\fi

\newcommand\appendixref{Appendix\xspace}

\ifextras
\newcommand\extraref{Appendix\xspace}
\newcommand\extra[1]{#1}
\else
\newcommand\extraref{full version of this paper~\cite{checkdpreport}}
\newcommand\extra[1]{}
\fi

\usepackage[utf8]{inputenc}
\usepackage[T1]{fontenc}
\usepackage{microtype}
\usepackage{amsmath,amsthm,commath}
\usepackage{mathtools}
\usepackage[linesnumbered,lined,ruled,vlined]{algorithm2e}
\usepackage{paralist}
\usepackage{listings}
\usepackage{mathpartir} 
\usepackage{stmaryrd}
\SetSymbolFont{stmry}{bold}{U}{stmry}{m}{n}
\usepackage{xspace}
\usepackage{dsfont}
\usepackage{bbold}
\usepackage{setspace}
\usepackage{todonotes}
\usepackage{balance}
\usepackage{subcaption}
\usepackage{comment}
\usepackage{xr}
\usepackage[normalem]{ulem}
\usepackage{enumitem}
\usepackage{makecell}
\usepackage{multirow}

\setlist[enumerate]{leftmargin=0.5cm,itemsep=0cm,topsep=0.1em,parsep=0.1em}
\setlist[itemize]{leftmargin=0.5cm,itemsep=0cm,topsep=0.1em,parsep=0.1em}

\graphicspath{ {figures/} }

\lstdefinelanguage{z3}{
  keywords={declare, const, assert, Int, Real, and, or, forall, minimize, maximize, check, sat, get, model},
}

\lstdefinelanguage{WHILE}{
  keywords={Lap, while, if, then, else, cons, havoc},
}

\newcommand\appendixalgsize\footnotesize

\makeatother

\newtheorem{definition}{Definition}
\newtheorem{theorem}{Theorem}

\newcommand\algrule[1][.5pt]{\par\vskip.2\baselineskip\hrule height #1\par\vskip.2\baselineskip}
\newcommand\code[1]{\texttt{#1}}
\newcommand\pair[2]{\langle {#1}, {#2} \rangle}
\newcommand\laplace{\ensuremath{\textbf{\code{Lap}}}\xspace}
\newcommand\lapm[1]{\laplace~{#1}}
\newcommand\proves{\vdash}
\newcommand\bool{\code{bool}}
\newcommand\true{\code{true}}
\newcommand\false{\code{false}}
\newcommand\basety{\mathcal{B}}
\newcommand\real{{r}}
\newcommand\bexpr{{\mathbb{b}}}
\newcommand\dexpr{{\mathbb{d}}}
\newcommand\nexpr{{\mathbb{n}}}
\newcommand\ternary[3]{#1\mathbin{?}#2\mathbin{:}#3}
\newcommand{\diff}[1]{{\setlength{\fboxsep}{0.5pt}\colorbox{lightgray}{\ensuremath{#1}}}}

\newcommand\skipcmd{\code{\textbf{skip}}}
\newcommand\transform{\rightharpoonup}
\newcommand\tylist{\code{list}}
\newcommand\tyreal{\code{num}}
\newcommand\ruleref[1]{{\sc (#1)}}

\newcommand\assert[1]{\code{\textbf{assert}}({#1})}
\newcommand\ifcmd[3]{\code{\textbf{if}}~{#1}~\code{\textbf{then}}~{#2}~\code{\textbf{else}}~{#3}}
\newcommand\whilecmd[2]{\code{\textbf{while}}~{#1}~\code{\textbf{do}}~{#2}}
\newcommand\outcmd[1]{\code{\textbf{return}}~#1}
\newcommand\instrument[1]{\underline{#1}}
\newcommand\annotation[1]{\colorbox{lightgray}{\ensuremath{#1}}}
\newcommand\tT{{T_\eta}}
\newcommand\distance[1]{\ensuremath{\widehat{\text{#1}}}}
\newcommand\vpriv[1]{{\mathbf{v}_{{#1}}}}
\newcommand\priv{\ensuremath{\epsilon}}
\newcommand\configtwo[2]{\langle {#1},~{#2} \rangle}
\newcommand\configthree[3]{\langle {#1},{#2},{#3} \rangle}
\newcommand\alignd{\circ}
\newcommand\shadow{\dagger}
\newcommand\first[1]{#1^\alignd}
\newcommand\second[1]{#1^\shadow}
\newcommand\third[1]{#1^\star}
\newcommand\select{\mathcal{S}}
\newcommand\funsigfour[4]{
\textbf{function}~\textsc{#1}~(#2) \par
\textbf{returns}~{#3}\par
\textbf{precondition}~{#4}\par}
\newcommand\funsigthree[3]{
\textbf{function}~\textsc{#1}~(#2) \par
\textbf{returns}~{#3}\par}
\newcommand\alldiffer{$\forall$ \code{i}.~ $-1\leq$~\code{\distance{q}[i]}~$\leq1$}
\newcommand\onediffer{$\forall\!$ \code{i}.~ $-1\!\leq\!$~\code{\distance{q}[i]}~$\!\leq\!1$ $\!\land\!$
($\forall\!$ \code{i}.~(\code{\distance{q}[i]}$\neq 0$) $\Rightarrow$ ($\forall\!$ \code{j}.~\code{\distance{q}[j] = 0}))}
\newcommand\alignment{\ensuremath{\mathcal{A}}\xspace}
\newcommand\exprset{\ensuremath{\mathbb{E}}\xspace}
\newcommand\varset{\ensuremath{\mathbb{V}}\xspace}
\newcommand\shadowexec[1]{\llparenthesis {#1} \rrparenthesis^\dagger}
\newcommand\aligndexec[1]{\llparenthesis {#1} \rrparenthesis^\circ}
\newcommand\op{\code{op}}

\newcommand\nvars{\text{\em V}}
\newcommand\rvars{\text{\em H}}

\newcommand\algoinput{\ensuremath{\mathcal{D}}\xspace}
\newcommand\algooutput{\ensuremath{\mathcal{O}}\xspace}
\newcommand\mechanism{\ensuremath{M}\xspace}

\newcommand\tool{CheckDP\xspace}
\newcommand\hole{\theta\xspace}

\newcommand\bound[1]{\textbf{check}(#1)}

\newcommand\unitop{\code{unit}}
\newcommand\bindop{\code{bind}}
\newcommand\dgdist{\ensuremath{\mathds{1}}\xspace}
\newcommand\dist{\mathbf{Dist}}
\newcommand\defn{\triangleq}

\newcommand\store{\ensuremath{m}}
\newcommand\Store{\ensuremath{\mathcal{M}}}
\newcommand\subst[3]{{#1}\{{#3}/{#2}\}}

\newcommand\join{\sqcup}

\newcommand\env{\Gamma}
\newcommand\flowrule[4]{#1~\{#2\transform#3\}~#4}
\newcommand\exprrule[4]{#1:#2 \mid #4}
\newcommand\noconstraints{\true}
\newcommand\constraints{\mathcal{C}\xspace}
\newcommand\noise{sample}

\newcommand\pc{pc}

\newcommand{\prob}{\ensuremath{\operatorname{\mathbb{P}}}} %
\newcommand{\mech}{\mechanism} %


\title{\tool: An Automated and Integrated Approach for Proving Differential Privacy or Finding Precise Counterexamples}

\author{Yuxin Wang, Zeyu Ding, Daniel Kifer, Danfeng Zhang}
\affiliation{
  \institution{The Pennsylvania State University}
}
\email{{yxwang,zyding}@psu.edu, {dkifer,zhang}@cse.psu.edu}

\begin{document}
\fancyhead{}
\fancyfoot[C]{\thepage}
\begin{abstract}
We propose CheckDP, an automated and integrated approach for
proving or disproving claims that a mechanism  is differentially private.
CheckDP can find counterexamples for mechanisms with subtle bugs for which prior counterexample generators have failed. Furthermore, it was able to \emph{automatically} generate proofs for correct mechanisms for which no formal verification was reported before.
CheckDP is built on static program analysis,
allowing it to be more efficient and precise in catching infrequent events than  sampling based
counterexample generators (which run mechanisms hundreds of thousands of times to
estimate their output distribution). Moreover, its sound approach also allows 
automatic verification of correct mechanisms. When evaluated 
on standard benchmarks and newer privacy mechanisms, CheckDP generates proofs (for correct
mechanisms) and counterexamples (for incorrect mechanisms) within 70 seconds without any
false positives or false negatives. %
\end{abstract}

\maketitle

\section{Introduction}\label{sec:introduction}
Differential privacy~\cite{dwork06Calibrating} has been adopted in major 
data sharing initiatives by organizations  
such as Google~\cite{rappor,prochlo}, Apple~\cite{applediffp},
Microsoft~\cite{DingKY17}, Uber~\cite{elasticsensitivity} and the U.S. Census
Bureau~\cite{ashwin08:map,onthemap,Haney:2017:UCF,abowd18kdd}. 
It allows these organizations to collect and share data with provable bounds
on the information that is leaked about any individual.

Crucial to any differentially private system is the correctness of \emph{privacy mechanisms},
the underlying privacy primitives in larger privacy-preserving algorithms.
Manually developing the necessary rigorous proofs that a mechanism correctly protects privacy is a subtle and error-prone
process. For example, detailed explanations of significant errors in peer-reviewed 
papers and systems can be found in ~\cite{frankblog,ninghuisparse,ashwinsparse}. Such mistakes have led
to research in the application of formal verification for \emph{proving} that mechanisms satisfy differential privacy \cite{lightdp, shadowdp,
Aws:synthesis, Barthe12,EasyCrypt,BartheICALP2013,Barthe16,BartheCCS16}.
However, if a mechanism has a bug making its privacy claim incorrect, these techniques cannot \emph{disprove} the privacy claims -- a counterexample detector must be used instead \cite{Ding2018CCS,Bichsel2018CCS,diffproptest}. Finding a counterexample is typically a two-phase process that (1) first searches an infinitely large space for candidate counterexamples and then (2) uses an exact symbolic probabilistic solver like PSI \cite{psisolver} to verify that the counterexample is indeed valid. The search phase currently presents the most problems (i.e., large runtimes or failure to find counterexamples are most often attributed to the search phase).
\hl{Earlier} search techniques were based on 
 sampling (running a mechanism hundreds of thousands of times), which made them slow and inherently imprecise: even with enormous amounts of samples, they can still fail if a privacy-violating section of
code is not executed frequently enough or if the actual privacy cost  is slightly higher than the privacy claim. \hl{Recently, static program analyses were proposed to accomplish both goals~\cite{barthe2020, farina2020coupled}. However, they either only analyze a non-trivial but restricted class of programs~\cite{barthe2020}, or rely on heuristic strategies whose effectiveness on many sutble mechanisms is unclear~\cite{farina2020coupled}.}

In this paper, we present \tool, an automated and integrated tool for proving or disproving the
correctness of a mechanism that claims to be differentially private. Significantly, \tool automatically finds counterexamples via static analysis,
\hl{making it unnecessary to run the mechanism.} 
Like prior work \cite{Bichsel2018CCS}, \tool still uses PSI \cite{psisolver} at the end. However, replacing sampling-based search with  static analysis  enables \tool to find violations in a few seconds, while previous sampling-based methods~\cite{Ding2018CCS,Bichsel2018CCS} may fail even after running for hours. Furthermore, sampling-based methods may still require manual setting of some program inputs (e.g., DP-Finder~\cite{Bichsel2018CCS} requires additional arguments to be set manually for Sparse Vector Technique in our evaluation) while \tool is fully automated.
\hl{Furthermore, the integrated approach of \tool allows it to efficiently analyze a larger class of differentially privacy mechanisms, compared with concurrent work using static analyses~\cite{barthe2020, farina2020coupled}.}

Meanwhile, \tool still offers state-of-the-art verification capability compared with existing language-based verifiers and is further able to automatically generate proofs for 3 mechanisms for which no formal verification was reported before. 
\tool takes the source code of a mechanism along with its claimed level of privacy and either generates a proof of correctness or a verifiable counterexample (a pair of related inputs 
and a feasible output).
\tool is built upon a proof technique called
\emph{randomness alignment}~\cite{lightdp, shadowdp, freegap}, which
recasts the task of proving differential privacy into one of finding \emph{alignments}
between random variables used by two related runs of the mechanism. 
\tool 
uses a novel verify-invalidate loop that alternatively improves tentative
proofs (in the form of alignments), which are then used to improve tentative counterexamples (and vice versa) until either the
tentative proof has no counterexample, or the tentative counterexample has no alignment. 

We evaluated \tool on correct/incorrect versions of existing benchmarks and newly proposed mechanisms.
It  generated a proof for each
correct mechanism within 70 seconds and a counterexample for each incorrect
mechanism within 15 seconds.

In summary, this paper makes the following contributions:
\begin{enumerate}
\item  \tool, \hl{one of} the first automated tools \hl{(with concurrent work \cite{barthe2020, farina2020coupled})} that generates both proofs for correct
mechanisms and counterexamples for incorrect mechanisms
(Section~\ref{sec:overview}).

\item A syntax-directed translation from the probabilistic mechanism being
checked to non-probabilistic target code with explicit proof obligations
(Section~\ref{sec:translation}).

\item An alignment template generation algorithm
(Section~\ref{sec:templates}).

\item A novel verify-invalidate loop that incrementally improves tentative proofs and counterexamples (Section~\ref{sec:tool}).

\item Case studies and experimental comparisons between \tool and  existing tools using correct/incorrect versions of existing benchmarks and newly proposed mechanisms. For incorrect mechanisms, \tool automatically found counterexamples in all cases, even in cases where competing methods \cite{Ding2018CCS,Bichsel2018CCS} failed. For correct mechanisms, \tool automatically generated proofs of privacy, including proofs for 3 mechanisms for which no formal verification was reported before (Section~\ref{sec:impexp}).

\end{enumerate}

\section{Preliminaries and Running Example
}\label{sec:background}

\subsection{Differential Privacy}\label{sec:diffpriv}

Among several popular variants of differential privacy
\cite{dwork06Calibrating,dworkKMM06:ourdata,BS2016:zcdp,M2017:Renyi}, we focus on \emph{pure} differential privacy~\cite{dwork06Calibrating}.
The goal of differential privacy is to hide the effect of any person's record
on the output of an algorithm. 
This is achieved by considering all pairs of datasets $D$ and $D^\prime$ that differ on one record.
We call such datasets \emph{adjacent} and denote it by $D\sim D'$.
To offer
privacy, a differentially private algorithm injects carefully calibrated random
noise during its computation. Given a pair of datasets $(D, D^\prime)$, we call the execution of an algorithm on $D$ 
the  \emph{original execution} and the execution on (neighboring) $D^\prime$ the \emph{related execution}.
Intuitively, we say a randomized algorithm is differentially private if the
output distribution of the original execution and its related execution are
hard to distinguish for all such dataset pairs:

\begin{definition}[Pure Differential Privacy \cite{Dwork06diffpriv}]
    \label{def:diffpriv}
    Let $\epsilon \geq 0$. A probabilistic computation $\mechanism: \algoinput \rightarrow \algooutput$ is $\priv$-differentially private  if for every pair of neighboring datasets $D\sim D' \in \algoinput$ and
    every output $o \in \algooutput$, 
      $ \prob[M (D) = o]\leq e^\priv \prob[M (D') = o]. $%
\end{definition}  

Often, a differentially private algorithm $M$ interacts with a dataset $D$
through a list of queries $f_1, f_2, \ldots$: it iteratively runs a query $f_i$
on $D$ to get an exact answer $q_i$, then performs some randomized computation
on the set of query answers $\{q_j \mid j \leq i\}$. We call the vector
$(q_1,q_2,\ldots)$ along \hl{with} other \hl{data-independent} parameters to $M$ (e.g.,
privacy parameter $\epsilon$) an \emph{input} to $M$.  The notion of adjacent
 datasets  translates into the notion of \emph{sensitivity} on those
queries:

\begin{definition}[Global Sensitivity \cite{diffpbook}]
    \label{def:sensitivity}
    The global sensitivity of a query $f$ is
$\Delta_{f} = \sup_{D\sim D'} \abs{f(D)-f(D')}$.
\end{definition} 

We say two inputs $inp = \{(q_1,q_2,\ldots), \text{params}\}$ and $inp'
= \{(q_1',q_2',\ldots),\text{params}\}$  are adjacent with respect to the queries $f_1, f_2, \ldots$, and write $inp\sim
inp'$, if \hl{the params are the same and } there exist two adjacent datasets $D$ and $D'$ such
that $(f_1(D), f_2(D), \dots) = (q_1, q_2,\ldots)$ and $(f_1(D'), f_2(D'),
\dots) = (q_1', q_2',\ldots)$. Note that this implies that
$\abs{q_i-q'_i}\leq \Delta_{f_i}, \forall i$. It  follows that
differential privacy can be proved by showing that for all pair of inputs
$inp\sim inp'$ and all outputs $o\in \algooutput$,
$\prob[M(inp) = o] \leq e^\priv \prob[M(inp') = o]$.
As standard, we assume that the sensitivity of inputs are either manually
specified or computed by sensitivity analysis tools (e.g.,~\cite{Fuzz,DFuzz}).

Many mechanisms are built on top of the Laplace Mechanism \cite{dwork06Calibrating} which adds Laplace noise to query answers:
\begin{theorem}[Laplace Mechanism \cite{dwork06Calibrating}]
Let $\epsilon>0$, let $D$ be a dataset, let $f$ be a query with sensitivity $\Delta_f$ and let $q=f(D)$. The
Laplace Mechanism which, on input $q$, outputs $q + \eta$ (where $\eta$ is
sampled from the Laplace distribution with mean 0 and scale parameter
$\Delta_{f}/\epsilon$) satisfies $\epsilon$-differential privacy. 
\end{theorem}
We sometimes abuse notation and refer to the sensitivity $\Delta_q$ of a numerical value $q$ -- we always take this to mean as the sensitivity of the function that produced $q$.

\subsection{Randomness Alignment}\label{sec:randomness_aligment}
\emph{Randomness alignment} is a simple yet powerful proof technique that underpins the verification tools LightDP~\cite{lightdp} and its successor
ShadowDP~\cite{shadowdp}. Precise reasoning using this proof technique was  used to improve a variety of algorithms, allowing them to release strictly
more information at the same privacy cost~\cite{freegap}. Given two executions of a randomized algorithm $M$ on $D$ and $D^\prime$ respectively,
a randomness alignment is a mapping between the random variables in the first execution to random variables in the second execution that will cause the second
execution to always produce the same output as the first. Upper bounds on privacy parameters depend on how much the random variables change under this mapping~\cite{lightdp}.

We use the Laplace Mechanism \cite{diffpbook} to illustrate the key ideas
behind randomness alignment.  Let $D\sim D'$ be a pair of neighboring datasets and
let $f$ be a query with sensitivity $\Delta_f$. Let $q=f(D)$ and $q'=f(D')$ be the respective query answers. %
If we use the Laplace Mechanism to answer these queries with privacy, on input $q$ (resp. $q'$) it will output $q+\eta$ (resp. $q'+\eta'$) 
where $\eta$ (resp. $\eta'$) is a Laplace random variable with scale $\Delta_f/\epsilon$.
In order for the Laplace Mechanism to produce the same output in both executions, we
need $q+\eta = q'+\eta'$ and therefore $\eta'=\eta + q - q'$. This creates a
``mapping'' between the values of random noises: if we change the input from
$q$ to $q'$, we need to adjust the random noise by an amount of $q-q'$ (i.e., this is the \emph{distance} we need to move $\eta'$ to get to $\eta$). Clearly
$\abs{q-q'}\leq \Delta_{f}$ by definition of sensitivity. The privacy proof follows
from the fact that if two random samples $\eta$ and $\eta'$ (from the
Laplace distribution with scale $\Delta_{f}/\epsilon$) are at most distance $\Delta_{f}$ apart,  
the ratio of their probabilities is at most $e^\epsilon$. Hence, the \emph{privacy cost}, the natural log of this ratio, is bounded by $\epsilon$.

Thus randomness alignment can be viewed in terms of \emph{distances} that we need to move random
variables.
Let $q \sim q'$ be query answers from neighboring datasets and $M$ be a randomized algorithm which uses a set of
random noises $H=\set{\eta}$. We associate to every random variable $\eta$ a numeric value
\distance{$\eta$} which tracks precisely the amount in value we need to change
$\eta$ in order to obtain the same output when the input to $M$ is changed from
$q$ to $q'$. In other words, the output of $M$ with input $q$ and random values
$\set{\eta}$ is the same as that of $M$ with input $q'$ and random values
$\set{\eta + \distance{$\eta$}}$. Taking $M$ to be the Laplace Mechanism, then
the alignment in the previous paragraph is
$\set{\distance{$\eta$} = q - q'}$. Note that the alignment is a function that depends on $M$ as
well as $q$ and $q'$.

If all of the random variables are Laplace, the cost of an alignment is the summation of $\frac{\text{distance}}{\text{noise scale}}$ for each random variable. To find the overall privacy cost (e.g., the $\epsilon$ in differential privacy), we then find an upper bound on the alignment cost for all related $q$ and $q'$.

\subsection{Privacy Proof and Counterexample}\label{sec:proof_and_counterexample}
Not all randomness alignments serve as proofs of differential privacy. To form
a proof, one must show that (1) the alignment forces the two related executions to produce the same output, (2) the privacy cost of an
alignment must be bounded by the promised level of privacy, and (3) the alignment is injective.
Hence, in this paper, an (alignment-based) privacy proof refers to a randomness alignment that
satisfies these requirements.

On the other hand, to show that an algorithm violates differential privacy, it
suffices to demonstrate the existence of a counterexample. Formally, if an
algorithm $\mech$ claims to satisfy $\epsilon$-differential privacy, a
\emph{counterexample} to this claim is a triple $(inp,inp',o)$ such that
$inp\sim inp'$ and $\prob[\mech(inp) = o] > e^\priv \prob[\mech(inp') = o]$.

\paragraph*{Challenges}
LightDP~\cite{lightdp} and ShadowDP~\cite{shadowdp} can check if a manually generated alignment is an alignment-based privacy proof. On the other hand, an exact symbolic probabilistic solver, such as PSI~\cite{psisolver},
can check if a counterexample, either generated manually or via a sampling-based generator, witnesses violation of differential privacy. To the best of our knowledge, \tool is the first tool that \emph{automatically generates} alignment-based proofs/counterexamples via static program analysis.\footnote{Prior work~\cite{Aws:synthesis} automatically generates coupling proofs, an alternative language-based proof technique for differential privacy. But all existing verifiers using alignment-based proofs\cite{lightdp,shadowdp} require manually provided alignments.}  To do so, \emph{a key challenge} is to
tackle the infinite search space of proofs
(i.e., alignments) and counterexamples. \tool uses a novel proof template  generation algorithm to reduce the search space of candidate alignments
(Section~\ref{sec:translation}) and uses a novel verify-invalidate loop
(Section~\ref{sec:tool}) to find tentative proofs, counterexamples
showing their privacy cost is too high, improved proofs, improved counterexamples, etc.

\subsection{Running Examples}\label{sec:overview}
To illustrate our approach, we now discuss two variants of
the Sparse Vector Technique~\cite{diffpbook}, one correct and one incorrect.
Using the two variants, we sketch how \tool automatically proves/disproves  (as appropriate) their claimed privacy properties.

\paragraph*{Sparse Vector Technique (SVT)~\cite{diffpbook}} A powerful mechanism proven
to satisfy differential privacy. It can be used as a building block for many
advanced differentially private algorithms. This mechanism is designed to solve
the following problem: given a series of queries and a preset public threshold,
we want to identify the first $N$ queries whose answers are above the threshold,
but in a privacy-preserving manner. To achieve this, it adds independent
Laplace noise both to the threshold and each query answer, then it returns the identities of the
first $N$ queries whose \emph{noisy} answers are above the noisy threshold. The
standard implementation of SVT outputs $\true$ for the above-threshold queries
and $\false$ for the others (and terminates when there are a total of $N$ outputs equal to $\true$). We use two variants of SVT for an overview of
\tool.

\paragraph*{GapSVT} This is an improved (and correct) variant of SVT which provides numerical information about
some queries. When a noisy query exceeds the noisy threshold, it outputs the difference between these noisy values; otherwise it returns $\false$. This provides an estimate for how much higher a query is compared to the threshold.
 The algorithm was first proposed and verified in~\cite{shadowdp}; its pseudo code is shown in
Figure~\ref{alg:gapsvt}. Here, $\lapm{(2/\priv)}$ draws one sample from Laplace
distribution with mean 0 and scale factor of $2/\priv$. This random value is
then added to the public threshold $T$ (stored as noisy threshold $\tT$). For
each query answer, another independent Laplace noise $\eta_2 =
\lapm{(4N/\priv)}$ is added. If the noisy query answer \code{q[i] + }$\eta_2$
is above the noisy threshold $\tT$, the gap between them (\code{q[i] + }$\eta_2
- \tT$) is added to the output list \code{out}, otherwise \code{0} is added.

\begin{figure}[ht]
\setstretch{0.9}
\raggedright
\small
\noindent\rule{\linewidth}{0.8pt}
\funsigfour{GapSVT}
{\code{T,N,size}\annotation{:\tyreal_{0}},\code{q}\annotation{:\tylist~\tyreal_{*}}}
{(\code{out}\annotation{:\tylist~\tyreal_{0}}), \bound{\priv}}
{\alldiffer}
\algrule
\begin{lstlisting}[frame=none,escapechar=@]
$\eta_1$ := $\lapm{(2/\priv)}$@\label{line:gapsvt_eta_1}@
$\tT$ := $T + \eta_1$;
count := 0; i := 0;
while (count < N $\land$ i < size)@\label{line:gapsvt_while}@
  $\eta_2$ := $\lapm{(4N/\priv)}$@\label{line:gapsvt_eta2}@
  if ($\text{q[i]}+\eta_2\geq \tT$) then
    out := (q[i] + $\eta_2$ - $\tT$)::out;@\label{line:gapsvt_output}@
    count := count + 1;
  else
    out := false::out;
  i := i + 1;
\end{lstlisting}

\noindent\rule{\linewidth}{0.8pt}
\funsigthree{Transformed GapSVT}
{\code{T,N,size,q}\instrument{, \code{\distance{q}}, $\noise$, $\hole$}}
{(\code{out})}
\algrule
\begin{lstlisting}[frame=none,escapechar=@,firstnumber=12]
@\instrument{$\vpriv{\priv}$ := 0; idx = 0;}@
@\instrument{$\eta_1$ := $\noise$[idx]; idx := idx + 1;}@
@\instrument{$\vpriv{\epsilon}$ := $\vpriv{\epsilon}$ + $|\alignment_1| \times \priv / 2$; \distance{$\eta_1$} := $\alignment_1$;}@
$\tT$ := $T$ + $\eta_1$;
@\instrument{\distance{$\tT$} := \distance{$\eta_1$};}@@\label{line:gapsvt_update_distance_tT}@
count := 0; i := 0;
while (count < N $\land$ i < size)
  @\instrument{$\eta_2$ := $\noise$[idx]; idx := idx + 1;}@
  @\instrument{$\vpriv{\priv}$ := $\vpriv{\priv}$ + $|\alignment_2| \times \priv / 4N$; \distance{$\eta_2$} := $\alignment_2$;}\label{line:gapsvt_eta2_cost}@
  if (q[i] + $\eta_2$ $\geq$ $\tT$) then
    @\instrument{\assert{q[i] + $\eta_2$ + \distance{q}[i] + \distance{$\eta_2$} $\geq$ $\tT$ + \distance{$\tT$}};}@@\label{line:gapsvt_true_assertion}@
    @\instrument{\assert{\distance{q}[i] + \distance{$\eta_2$} - \distance{$\tT$} == 0};}@
    out := (q[i] + $\eta_2$ - $\tT$)::out;
    count := count + 1;
  else
    @\instrument{\assert{$\lnot$(q[i] + $\eta_2$ + \distance{q}[i] + \distance{$\eta_2$} $\ge$ $\tT$ + \distance{$\tT$})};}@@\label{line:gapsvt_false_assertion}@
    out := false::out;
  i := i + 1;
@\instrument{\assert{$\vpriv{\priv} \le \priv$};}@
\end{lstlisting}
\noindent\rule{\linewidth}{0.8pt}
\vspace{-5ex}
\caption{GapSVT and its transformed code, where underlined parts are added by \tool. The transformed
code contains two alignment templates for $\eta_1$ and $\eta_2$:
$\alignment_1 = \hole[0]$ and $\alignment_2 = (\code{$q[i] +
\eta_2$ $\ge$ $\tT$})\mathbin{?}(\hole[1] + \hole[2] \times
\distance{$\tT$} + \hole[3] \times
\distance{\code{$q$}}\code{$[i]$})\mathbin{:}(\hole[4] + \hole[5] \times \tT +
\hole[6] \times \distance{\code{$q$}}\code{$[i]$})$. The random variables
and $\hole$ are inserted as part of the function input.\label{alg:gapsvt}}
\vspace{-3ex}
\end{figure}

One key observation from the manual proofs of SVT and its
variants~\cite{diffpbook,ninghuisparse,ashwinsparse,freegap} is that the privacy cost
is only paid for the queries whose noisy answers are above the noisy threshold. In other words, outputting $\false$ does not
incur any new privacy cost. Correspondingly, the correct alignment
for GapSVT~\cite{shadowdp,freegap} (that is, the  distance that $\eta_1$ and $\eta_2$ need to be moved to ensure the output is the same when the input changes from $q[i]$ to $q'[i]\equiv q[i]+\distance{$q$}[i]$, for all $i$)  is:
$
\eta_1: 1$ and $ \eta_2: \code{q[i] + $\eta_2$ $\ge$ $\tT$}\mathbin{?}\code{(1 - \distance{q}[i])}\mathbin{:}0
$.

Note that $\eta_2$ is aligned with non-zero distance only under the $\true$
branch; hence, no privacy cost is paid in the other branch. It is easy to verify that
if every query has sensitivity 1, the cost of this alignment is bounded by $\epsilon$.

\paragraph*{BadGapSVT}
We also consider a variant of SVT (and GapSVT) that incorrectly tries to release numerical information.
When a noisy query answer is larger than the noisy threshold, the variant releases that noisy query answer (that is, it \emph{does not} subtract from it the noisy threshold); otherwise it outputs $\false$.  This is
an incorrect variant of SVT~\cite{aaronnotes} that was reported
in~\cite{ninghuisparse} and was called iSVT4 in~\cite{Ding2018CCS}.  
More precisely, BadGapSVT replaces
line~\ref{line:gapsvt_output} of GapSVT with \code{out := (q[i]
+ $\eta_2$)::out;}. This small change makes it not $\epsilon$-differentially private \cite{ninghuisparse}.
The reason why is subtle, but the intuition is the following.
Suppose BadGapSVT returns a noisy query answer
\code{q[i]} + $\eta_2$ = $3$, the attacker is able to deduce that $\tT \le
3$. Once this information is leaked,  outputting $\false$ in the else branch 
is no longer ``free''; every output incurs a privacy cost. %

\subsection{Approach Overview}\label{sec:approach_overview}

We use GapSVT and BadGapSVT to illustrate how \tool  generates
proofs and counterexamples.

\paragraph*{Code Transformation (Section~\ref{sec:translation})} 
\tool first takes the probabilistic algorithm being checked, written in the \tool
language (Section~\ref{sec:syntax}), and generates the non-probabilistic target
code with \emph{assertions} and \emph{alignment templates} (i.e. templates for possible alignments).  The bottom of
Figure~\ref{alg:gapsvt} shows the transformed code of GapSVT with alignment
templates. The transformed code is distinguished from the source code in a few
important ways:
\begin{inparaenum}[(1)] 
\item The probabilistic sampling commands (at lines 1 and 5) are replaced by
non-probabilistic counterparts that read samples from the instrumented function
input $sample$.
\item An alignment template (e.g., $\alignment_1$, $\alignment_2$) is generated
for each sampling command; each template contains a few
holes, i.e., $\hole$, which is also instrumented as function input.
\item A distinguished variable $\vpriv{\priv}$ is added to track the overall
privacy cost and lines 14 and 20 update
the cost variable in a sound way.
\item 
Assertions are inserted in the transformed code
(lines~\ref{line:gapsvt_true_assertion},23,\ref{line:gapsvt_false_assertion},30)
to ensure the following soundness property:
\vspace{-2ex}
\begin{multline*}
\text{if $M(inp)$ is transformed to
$M'(inp,\distance{$inp$},sample,\hole)$, then }\\
\exists \hole.~\forall inp,\distance{$inp$},sample.~\textit{all assertions in $M'$ pass}\\
\implies M \text{ is differentially private}
\end{multline*}
\vspace{-4ex}
\end{inparaenum}

We note that the transformed code forms the basis for both proof and counterexample generation in \tool.

\paragraph*{Proof/Counterexample Generation (Section~\ref{sec:tool})}
 
Inspired by the Counterexample Guided Inductive Synthesis (CEGIS)~\cite{CEGIS} technique, originally proposed for program synthesis, \tool uses a verify-invalidate loop to simultaneously generate proofs and counterexamples. Unlike CEGIS, however,
the verify-invalidate loop is \emph{bidirectional}, in the sense that it internally records 
all previous counterexamples (resp. proofs) to generate one proof (resp. counterexample) as the algorithm output. On the other hand, the CEGIS loop is \emph{unidirectional}: it only collects and uses a set of inputs to guide synthesis internally.
At a high level, the verify-invalidate loop of \tool includes two integrated sub-loops, one for proof generation and the other for counterexample generation.

\paragraph{Verify Sub-loop} Its goal is to generate a proof (i.e., an
instantiation of $\hole$) such that $$\forall inp,\distance{$inp$},sample.~\textit{all assertions in $M'$ pass}$$

This is done by two iterative phases:
\begin{enumerate}
\item Generating invalidating inputs: Given a proof candidate (i.e., an
instantiation of $\hole$), it is \emph{incorrect} if $$\exists inp,\distance{$inp$},sample.~\textit{some assertion in $M'$ fails}$$

We use $I$ to denote a triple of $inp,\distance{$inp$},sample$.  Hence, given any
instantiation of $\hole$, we use an off-the-shelf symbolic execution tool
such as KLEE~\cite{klee} to find invalidating inputs when possible.

\item Generating proof candidates: with a set of invalidating inputs found so
far $I_1, \cdots, I_i$, we can try to generate a new proof candidate to satisfy
$$\exists \hole.~M'(I_1,\hole)\land \cdots \land M'(I_i,\hole)$$

\end{enumerate}

Starting from a default instantiation (e.g., one that sets $\forall
i.~\hole[i]=0$), \tool iteratively repeats Phases 1 and 2. Since
\tool uses all invalidating inputs found so far in Phase 2, the
proof candidate after each iteration is improving. 
When Phase 1 gets stuck, \tool obtains a proof candidate $\hole$ which is a privacy proof if $$\forall inp,\distance{$inp$},\noise.~M'(inp,\distance{$inp$},sample,\hole)$$
due to the soundness
property above. Hence, a proof (alignment) can be validated by program verification tools
such as CPAChecker~\cite{beyer2011cpachecker}. For GapSVT, \tool generates and
verifies (via CPAChecker)
that $\hole=\{1,1,0,-1,0,0,0\}$ results in a proof that GapSVT satisfies
$\epsilon$-differential privacy.

\paragraph{Invalidate Sub-loop}
While the verify sub-loop is conceptually similar to a CEGIS loop~\cite{CEGIS}, \tool also employs an invalidate sub-loop (integrated with the verify sub-loop); its goal is to generate \emph{one invalidating input} $I$ such that $\forall \hole.~ \textit{some assertion in $M'$ fail}$.
This is done by two iterative phases:\footnote{Note that  a set of invalidating inputs $I_1, \cdots, I_i$, generated from Phase 2 of the verify sub-loop is not a counterexample candidate, since by definition, a differential privacy counterexample consists of only one invalidating input.}

 \begin{enumerate}
    \item Generating proof candidates: Given an invalidating input $I$, it is \emph{incorrect} if
    $\exists \hole.~M'(I, \hole)$. Hence, given any $I$, we can use KLEE~\cite{klee} to find an alignment when possible.
    \item Generating counterexamples: with a set of previously found alignments $\hole_1, \cdots, \hole_i$, we try to find a new invalidating input to satisfy $$\exists I. \lnot M'(I, \hole_1) \land \cdots \land \lnot M'(I, \hole_i)$$
\end{enumerate}   
To integrate with the verify sub-loop, Phase 1 of the invalidate sub-loop starts when Phase 2 of the verify sub-loop gets stuck with a set of invalidating inputs $I_1, \cdots, I_i$; it uses $I_i$ to proceed since it is the most promising one. When Phase 1 of invalidate sub-loop gets stuck, \tool obtains a counterexample candidate, which can be validated by PSI~\cite{psisolver} (this is necessary since a mechanism might be differentially private even if no alignment-based proof exists).

For example, the
counterexample found for BadGapSVT sets the threshold $T=0$, $N=1$ (max number of outputs equal to $\true$ before termination), 
neighboring inputs $q=[0,0,0,0,0]$ and $q'= [1,1,1,1,-1]$, and the following output to examine $[0,0,0,0,1]$. PSI confirms that
the probability of this output when $q$ is an input is $\geq e^\epsilon$ times the probability of this output when $q'$ is the input.

When Phase 1 of the invalidate sub-loop generates a new alignment $\hole$, which happens in our empirical study (Section~\ref{sec:impexp}), Phase 2 follows to generate an ``improved'' invalidating input, which is then used to start Phase 2 of the validate sub-loop.

\section{Program Transformation}\label{sec:translation}
\tool takes a probablistic program along with an adjacency specification (i.e.,
how much two adjacent inputs can differ) and the
claimed level of differential privacy as inputs. It
translates the source code into a non-probabilistic program
with assertions to ensure differential privacy.
The transformed code forms the basis of finding a proof or a counterexample (Section~\ref{sec:tool}).

\subsection{Syntax}\label{sec:syntax}
\begin{figure}
\setstretch{0.9}
$
\begin{array}{lccl}
\text{Reals} & r &\in &\mathbb{R} \\
\text{Booleans} & b &\in &\{\true,\false\} \\
\text{Vars} & x  &\in& \nvars \\
\text{Rand Vars} & \eta &\in& \rvars\\
\text{Linear Ops} & \oplus &::= &+ \mid - \\
\text{Other Ops} & \otimes &::= &\times \mid / \\
\text{Comparators} & \odot &::= &< \mid > \mid = \mid \leq \mid \geq \\
\text{Bool Exprs} & \bexpr &::=\; &\true \mid \false \mid x \mid \neg \bexpr \mid \nexpr_1 \odot \nexpr_2 \\
\text{Num Exprs} & \nexpr &::=\; &\real \mid x \mid \eta  \mid \nexpr_1\oplus \nexpr_2 \mid \nexpr_1\otimes \nexpr_2 \mid \bexpr\mathbin{?}\nexpr_1\mathbin{:}\nexpr_2 \\
\text{Expressions} & e &::=\; &\nexpr \mid \bexpr \mid e_1::e_2 \mid e_1[e_2] \\
\text{Commands} & c &::=\; &\skipcmd \mid x := e \mid \eta  := g \mid c_1;c_2 \mid  \\ & & & \ifcmd{e}{(c_1)}{(c_2)} \mid \\ & & & \whilecmd{e}{(c)} \mid \outcmd e\\
\text{Rand Exps}   & g &::=\; &\lapm{\real}\\
\text{Types} & \tau &::=\; &\tyreal_{\dexpr} \mid \bool \mid \tylist~\tau \\
\text{Distances}& \dexpr &::= &0 \mid *\\
\end{array}
$
\caption{\tool: language syntax.}\label{fig:syntax}
\end{figure}

The syntax of \tool source code is listed in Figure~\ref{fig:syntax}.  Most
of the syntax is standard with the following features: 
\begin{itemize}
    \item Real numbers, booleans and their standard operations;
    \item Ternary expressions $\mathbb{b}\mathbin{?}\mathbb{n}_1\mathbin{:}\mathbb{n}_2$, it returns $\mathbb{n}_1$ when $\mathbb{b}$ evaluates to $\true$ or $\mathbb{n}_2$ otherwise;
    \item List operations: $e_1 :: e_2$ appends element $e_1$ to list $e_2$, and $e_1[e_2]$ gets the $e_2^\text{th}$ element of list $e_1$;
    \item Loop with keyword \textbf{\code{while}} and branch with keyword \textbf{\code{if}};
    \item A final return command $\outcmd{e}$.
\end{itemize}  

We now introduce other interesting parts that are needed for developing
differentially private algorithms.

\paragraph*{Random Expressions}
Differential privacy relies heavily on probabilistic computations: many
mechanisms achieve differential privacy by adding appropriate random noise to
variables. To model this behavior, we embed a sampling command $\eta:=\lapm{r}$
in \tool, which draws a sample from the Laplace distribution with mean 0 and
scale of $r$. In this paper, we only focus on the most interesting sampling
command $\lapm{r}$ (which is used in Laplace Mechanism and GapSVT in
Section~\ref{sec:background}). However, 
we note that it is fairly easy to add new sampling distributions to \tool.

For clarity, we distinguish variables holding random values, denoted by $\eta \in \rvars$, from other ones, denoted by $x \in \nvars$. 

\paragraph*{Types with Distances}
To enable alignment-based proof, one important aspect of the type system in
\tool is the ability to compute and track the distances for
each program variable. Motivated by verification tools using alignments (e.g.,
LightDP~\cite{lightdp} and ShadowDP~\cite{shadowdp}), types in the source language of \tool have the
form of $\basety_0$ or $\basety_*$, where $\basety$ is the base
type such as numerics ($\tyreal$), booleans ($\bool$) and lists
($\tylist~\tau$). The subscript of each type is the key to  alignment-based
proofs: it explicitly tracks the \emph{exact} difference between the value of a
variable in two related runs.

In the source language of \tool, the distances can either be $0$ or $*$: the former
indicates the variables stay the same in the related runs; the latter
means that the variable might hold different values in two related runs and the
value difference is stored in a distinguished variable $\distance{$x$}$ added
by the program transformation (i.e., a syntactic sugar for dependent sum type
$\sum_{(\distance{$x$}:~\tyreal_0)}~\basety_{\distance{$x$}}$). For example,
inputs \code{T,N,size} are annotated with distance $0$ in
Figure~\ref{alg:gapsvt}, meaning that they are public parameters to the
algorithm; query answers $q$ are annotated with distance $*$,
meaning that each $q[i]$ differ by exactly $\distance{$q$}[i]$ in two related
runs. The type system distinguishes zero-distance variables as an
optimization: as we show shortly, it helps to reduce the code size for later
stages (Section~\ref{sec:commands}) as well as aids proof template generation
(Section~\ref{sec:templates}).

Note that boolean types ($\bool$) and list types ($\tylist~\tau$) cannot be
associated with numeric distances, hence omitted in the syntax. However, nested
cases such as $\tylist~{\tyreal_{*}}$ still accurately track the
distances of the elements inside the list.

\hl{The semantics of \tool follows the standard definitions of probabilistic programs~\cite{Kozen81}; the formal semantics can be found in the \extraref.} Finally, \tool also supports shadow execution, a technique that underpins ShadowDP~\cite{shadowdp} and is crucial to the verification of challenging mechanisms such as Report Noisy Max~\cite{Dwork06diffpriv}. However, in order to focus on the most interesting parts of \tool, we first present the transformation without shadow execution, and later discuss how to support it.

\move{
\subsection{Semantics}
Following the standard definition in~\cite{lightdp,shadowdp}, we define the denotational
semantics of the probabilistic language as a mapping from initial memory to a
distribution on a (possible) final outputs. More formally, let $\Store$ be a
set of memory states where each $\store \in \Store$ maps all variables to their
values, including normal variables ($x\in \nvars$ ) and random variables ($\eta \in
\rvars$ ). 

A program is interpreted as a function $\Store
\rightarrow \dist{(\basety)}$ where $\basety$ is the return type (i.e., the
type of $e$). Since the semantics is standard, we omit it in the paper.
}

\begin{figure*}
\small
\raggedright
\framebox{\textbf{Transformation rules for expressions with form $\env \proves e:\basety_{\nexpr}$}}
\begin{mathpar}
\inferrule*[right=(T-Num)]{ }
{ \env \proves \exprrule{\real}{\tyreal_0}{\real}{\noconstraints}}
\and
\inferrule*[right=(T-Boolean)]{ }
{\env \proves \exprrule{b}{\bool}{b}{\noconstraints}}
\and
\inferrule*[right=(T-VarZero)]{ }
{ \env, x:\basety_0 \proves \exprrule{x}{\basety_0}{x}{\noconstraints}}
\and
\inferrule*[right=(T-VarStar)]
{ }
{\env, x:\basety_{*}\proves \exprrule{x}{\basety_{\distance{$x$}}}{x + \distance{$x$}}{\noconstraints}}
\and
\inferrule*[right=(T-Neg)]
{\env\proves \exprrule{e}{\bool}{e'}{\constraints}}
{ \env \proves \exprrule{\neg e}{\bool}{e'}{\constraints}}
\and
\inferrule*[right=(T-OPlus)]
{\env\proves \exprrule{e_1}{\basety_{\nexpr_1}}{e_1'}{\constraints_1} \quad \env\proves \exprrule{e_2}{\basety_{\nexpr_2}}{e_2'}{\constraints_2}{}}
{\env \proves \exprrule{e_1 \oplus e_2}{\basety_{\nexpr_1 \oplus \nexpr_2}}{e_1' \oplus e_2'}{\constraints_1 \land \constraints_2}}
\and
\inferrule*[right=(T-OTimes)]
{\env\proves \exprrule{e_1}{\tyreal_{\nexpr_1}}{}{\constraints_1} \quad \env\proves \exprrule{e_2}{\tyreal_{\nexpr_2}}{}{\constraints_2}}
{\env \proves \exprrule{e_1 \otimes e_2}{\tyreal_0}{}{\constraints_1 \land \constraints_2 \land (\nexpr_1 = \nexpr_2 = 0)}}
\quad
\inferrule*[right=(T-ODot)]
{\env\proves \exprrule{e_1}{\tyreal_{\nexpr_1}}{}{\constraints_1} \quad \env\proves \exprrule{e_1}{\tyreal_{\nexpr_2}}{}{\constraints_2}}
{\env \proves \exprrule{e_1 \odot e_2}{\bool}{}{\constraints_1 \land \constraints_2 \land \inferrule{}{(e_1 \odot e_2) \Leftrightarrow \\\\ (e_1 + \nexpr_1) \odot (e_2 + \nexpr_2)}}}
\and
\inferrule*[right=(T-Cons)]
{\env\proves \exprrule{e_1}{\basety_{\nexpr_1}}{}{\constraints_1} \quad \env\proves \exprrule{e_2}{\tylist~\basety_{\nexpr_2}}{}{\constraints_2}}
{\env \proves \exprrule{e_1::e_2}{\tylist~\basety_{\nexpr}}{}{\constraints_1 \land \constraints_2 \land (\nexpr_1 = \nexpr_2 = 0)}}
\and
\inferrule*[right=(T-Index)]
{\env\proves \exprrule{e_1}{\tylist~\tau}{}{\constraints_1}\quad \env\proves \exprrule{e_2}{\tyreal_{\nexpr}}{}{\constraints_2}}
{\env \proves \exprrule{e_1[e_2]}{\tau}{}{\constraints_1 \land \constraints_2 \land (\nexpr = 0)}}
\and
\inferrule*[right=(T-Select)]
{\env\proves \exprrule{e_1}{\bool}{}{\constraints_1} \quad \env\proves \exprrule{e_2}{\basety_{\nexpr_1}}{}{\constraints_2} \quad \env\proves \exprrule{e_3}{\basety_{\nexpr_2}}{}{\constraints_3}}
{\env \proves \exprrule{e_1\mathbin{?}e_2\mathbin{:}e_3}{\basety_{\nexpr_1}}{}{\constraints_1 \land \constraints_2 \land \constraints_3\land (\nexpr_1=\nexpr_2)}}
\end{mathpar}
\framebox{\textbf{Transformation rules for commands with form $\proves \flowrule{\env}{c}{c'}{\env'}$}}
\begin{mathpar}

\inferrule*[right=(T-Asgn)]
{ \env \proves \exprrule{e}{\basety_{\nexpr}}{}{\constraints} \quad \configtwo{\dexpr}{c} = \begin{cases}
\configtwo{0}{\skipcmd} , \ \text{ if } \nexpr == 0, \cr
\configtwo{*}{\distance{$x$} := \nexpr} , \text{ otherwise} 
\end{cases}}
{\proves \flowrule{\env}{x := e;}{\assert{\constraints}; x := e; c}{\env[x\mapsto \basety_{\dexpr}]}}
\and
\inferrule*[right=(T-Seq)]
{\proves \flowrule{\env}{c_1}{c_1'}{\env_1} \quad \proves \flowrule{\env_1}{c_2}{c_2'}{\env_2}}
{\proves\flowrule{\env}{c_1;c_2}{c_1';c_2'}{\env_2}}
\and
\inferrule*[right=(T-Return)]
{ \env \proves \exprrule{e}{\basety_{\nexpr}}{}{\constraints}}
{\proves \flowrule{\env}{\outcmd{e}}{\assert{\constraints\land \nexpr = 0};\outcmd{e}}{\env}}
\and
\inferrule*[right=(T-Skip)]{ }
{\proves \flowrule{\env}{\skipcmd}{\skipcmd}{\env}}
\and
\inferrule*[right=(T-While)]
{\proves\flowrule{\env \join \env_f}{c}{c'}{\env_f} \quad \env, \env \join \env_f \Rrightarrow c_s \quad \env_f, \env \join \env_f \Rrightarrow c''
}
{\proves \flowrule{\env}{\whilecmd{e}{c}}{c_s; (\whilecmd{e}{(\assert{\aligndexec{e, \env}};c';c''))}}{\env \join \env_f}}
\and
\inferrule*[right=(T-If)]
{
\proves \flowrule{\env}{c_i}{c_i'}{\env_i} \quad \env_i, \env_1 \join \env_2 \Rrightarrow c''_i \quad i \in \{1, 2\} 
}
{\proves \flowrule{\env}{\ifcmd{e}{c_1}{c_2}}{\ifcmd{e}{(\assert{\aligndexec{e, \env}};c_1';c_1'')}{(\assert{\lnot\aligndexec{e, \env}};c_2';c_2'')}}{\env_1 \join \env_2}}
\and
\inferrule*[right=(T-Laplace)]
{\annotation{\alignment=\code{GenerateTemplate}(\Gamma,\text{All Assertions})}\quad c_a = \assert{(\subst{(\eta + \alignment)}{\eta}{\eta_1} = \subst{(\eta + \alignment)}{\eta}{\eta_2} \Rightarrow \eta_1=\eta_2)} } 
{\proves \flowrule{\env}{c_a; \eta := \lapm{\real}}{\eta := \noise[idx]; idx:= idx+1; \vpriv{\priv} := \vpriv{\priv}+|\annotation{\alignment}|/r;\distance{$\eta$} := \annotation{\alignment};}{\env[\eta \mapsto \tyreal_*]}}
\end{mathpar}
\framebox{\textbf{Transformation rules for merging environments}}
\begin{mathpar}
\inferrule*
{\env_1 \sqsubseteq \env_2 \quad c= \{\distance{$x$} := 0 \mid \env_1(x)=\tyreal_0 \land \env_2(x) = \tyreal_*\}}
{\env_1, \env_2 \Rrightarrow c}
\end{mathpar}
\caption{Program transformation rules. Distinguished variable $\vpriv{\priv}$ and assertions are added to ensure differential privacy.\label{fig:trans_rules}}
\end{figure*}

\subsection{Program Transformation}\label{subsec:transformation}

\tool is equipped with a flow-sensitive type system whose typing rules are shown in
Figure~\ref{fig:trans_rules}. At command level, each rule has the following
format: $\proves \flowrule{\env}{c}{c'}{\env'}$
where a typing environment $\env$ tracks for each program variable its type
with distance, $c$ and $c'$ are the source and target programs respectively,
and the flow-sensitive type system also updates typing environment to $\env'$
after command $c$. At a high-level, the type system
transforms the probabilistic source code $c$ into the
non-probabilistic target code $c'$ in a way that if all assertions in $c'$
holds, then $c$ is differentially private.

CheckDP's program transformation is motivated by those of LightDP
and ShadowDP~\cite{lightdp, shadowdp}, all built on randomness alignment proof. However,
there are a few important differences:
\begin{itemize}
\item \tool generates an alignment template for each sampling
instruction, rather than requiring manually provided alignments. 
\item \tool defers all privacy-related checks to assertions. This is crucial since
information needed for proof and counterexample generation is unavailable in a
lightweight static type system.
\item \tool only tracks if a variable has the same value in two
related runs (with distance 0) or not (with distance $*$).  This design aids
alignment template generation and reduces the size of transformed code.
\end{itemize}

\paragraph*{Checking Expressions}
Each typing rule for expression $e$ computes the correct distance for its
resulting value: $\env \proves \exprrule{e}{\basety_{\nexpr}}{}{\constraints}$, which reads as: expression $e$ has type $\basety$ and distance $\nexpr$ under
the typing environment $\env$ if the constraints $\constraints$ are satisfied. The
reason to collect constraints $C$ instead of statically checking them, is to
defer all privacy-related checks to later stages. %

Most of the expression rules are straightforward: they check the base types
(just like a traditional type system) and compute the distance of $e$'s value
in two related runs. For example, all constants must be identical
(Rules~\ruleref{T-Num,T-Boolean}) and the distance of a variable is retrieved
from the environment~\ruleref{T-VarZero,T-VarStar} (note that rule
\ruleref{T-VarStar} just desugers the $*$ notation). For linear operation
($\oplus$), the distance of the result is computed in a precise way
(Rule~\ruleref{T-OPlus}), while the other operations are treated in a more
conservative way: constraints are generated to ensure that the result is
identical in Rules~\ruleref{T-OTimes, T-ODot}. For example, \ruleref{T-ODot}
ensures boolean value of $e_1 \odot e_2$ will be the same in two related runs
by adding a constraint 
$$(e_1 \odot e_2) \Leftrightarrow (e_1 + \nexpr_1) \odot (e_2 + \nexpr_2)$$
\ruleref{T-Cons} restricts constructed list elements to have $0$-distance (note
that the restriction does not apply to input lists), while \ruleref{T-Index}
requires the index to have zero-distance.  Rule \ruleref{T-Select} restricts $e_1$
and $e_2$ to have the same distance.
The constraints gathered in the expression rules will later be explicitly
instrumented as assertions in the translated programs, which we will explain
shortly.

\subsection{Checking Commands}\label{sec:commands}
For each program statement, the type system updates the typing environment and if
necessary, instruments code to update $\distance{$x$}$ variables to the correct
distances. Moreover, it ensures that the two related runs take the same branch
in if-statement and while-statement.

\paragraph*{Flow-Sensitivity}
Each typing rule updates the typing environment to track if a variable has
zero-distance. When a variable has non-zero distance, it instruments the source
code to properly maintain the corresponding $\distance{$x$}$ variables.  The most
interesting rules are: rule~\ruleref{T-Asgn} properly promotes
the type of $x$ to be $\basety_{*}$ (tracked by distance variables) in $\env'$
if the distance of $e$ is not $0$. Meanwhile it optimizes away updates to $\distance{x}$ and properly downgrades
type to $\basety_0$ if $e$ has a zero-distance.
For example, line~\ref{line:gapsvt_update_distance_tT} in GapSVT
(Figure~\ref{alg:gapsvt}) is instrumented to update distance of $\tT$,
according to the distance of $T + \eta_1$. Moreover, variable \code{count} in
GapSVT always has the type $\tyreal_0$; therefore its distance variable never
appears in the translated program due to the optimization
in~\ruleref{T-Asgn}.

Rule~\ruleref{T-If} and \ruleref{T-While} are more complicated since they both
need to merge environments. In rule~\ruleref{T-If}, as $c_1$ and $c_2$ might
update $\env$ to $\env_1$ and $\env_2$ respectively, we need to merge them
in a natural way: the distance of a type form a two-level lattice
with $0\sqsubset *$. Thus we define a union operator $\join$ for distances $\dexpr$ as:
\[
\dexpr_1\join \dexpr_2 \defn
\begin{cases}
\dexpr_1 & \text{if } \dexpr_1 = \dexpr_2  \cr
* & \text{otherwise } \cr
\end{cases}
\]
therefore the union operator for two environments are defined as follows: $\env_1 \join \env_2 = \lambda x.~\env_1[x] \join \env_2[x]$.

Moreover, we use an auxiliary function $\env_1, \env_2 \Rrightarrow c$ to
``promote'' a variable to star type.
For example, with $\env(x)=*$, $\env(y)=*$ and $\env(b)=0$, rule~\ruleref{T-If}
translates the source code \\
$\ifcmd{b}{x:=y}{x:=1}$ to the following: \\
$\ifcmd{b}{(x:=y;\distance{$x$}:=\distance{$y$};)}{(x:=1;\distance{$x$}:=0)}$\\
where $\distance{$x$}:=\distance{$y$}$ is instrumented by~\ruleref{T-Asgn} and
$\distance{$x$}:=0$ is instrumented due to the promotion.

Similarly, the typing environments are merged in rule \ruleref{T-While}, except
that it requires a fixed point $\env_f$ such that $\proves \env\join\env_f\
\{c\}\ \env_f$. We follow the construction in~\cite{shadowdp} to compute a fixed point, noting that the
computation always terminates since all of the translation rules are monotonic
and the lattice only has two levels.

\paragraph*{Assertion Generation}

To ensure differential privacy, the type system inserts assertion in various
rules:
\begin{itemize}
\item To ensure that two related runs take the same control flow,
\ruleref{T-If} and \ruleref{T-While} asserts  that the value of the branch condition stays the same across two related executions. A helper function $\aligndexec{e, \env}$ is used to compute the value of $e$ in the aligned execution; its full definition can be found in the \appendixref.

\item To ensure that the final output value is differentially private,
rule~\ruleref{T-Return} asserts that its distance is zero (i.e., identical in
two related runs).

\item To ensure all constraints collected in the expression rules are
satisfied, assignment rules \ruleref{T-Asgn} and \ruleref{T-AsgnStar}
also insert corresponding assertions.
\end{itemize}

\subsection{Checking Sampling Commands}

Rule~\ruleref{T-Laplace} performs a few important tasks:

\paragraph*{Replacing Sampling Command} Rule \ruleref{T-Laplace}
removes the sampling instruction and assign to $\eta$ the next (unknown) sample
value $\code{\noise[idx]}$, where $\code{\noise}$ is a parameter of type
$\tylist~\tyreal$ added to the transformed code. The typing rule also
increments $\code{idx}$ so that the next sampling command will read out the
next value.

\paragraph*{Checking Injectivity}
T-Laplace adds an assertion $c_a$ to check the injectivity of the generated alignment (a fundamental requirement of alignment-based proofs): the same aligned value of $\eta$ implies the same value of $\eta$ in the original execution.

\paragraph*{Tracking Privacy Cost} A distinguished privacy cost variable
$\vpriv{\priv}$ is also instrumented to track the cost for aligning the random
variables in the program. Due to the properties of Laplace distribution, for a
sampling command $\eta := \lapm{r}$ with alignment template $\alignment$, we have $\prob
(\eta) / \prob (\eta + \alignment) \le e^{\abs{\alignment} / r}$. Hence, the privacy cost for
aligning $\eta$ by $\alignment$ is $\abs{\alignment} / r$. Note that the symbols in gray,
including $\alignment$, are placeholders when the rule is applied, since function
$\code{GenerateTemplate}$ takes all assertions in the transformed code as
inputs. Once translation is complete, the placeholders are filled in by the
algorithm that we discuss in Section~\ref{sec:tool}.

\paragraph*{Alignment Template Generation}\label{sec:templates}

For each sampling command $\eta := \lapm{r}$, an alignment of $\eta$ is needed in a randomness alignment proof. In its most flexible form, the alignment can be
written as any numerical expression $\nexpr$, which is prohibitive for our goal
of automatic proof generation. On the other hand, using simple heuristics such
as only considering constant alignment does not work: for example, the correct
alignment for $\eta_2$ in GapSVT is written as
``$(\code{q[i]} + \eta_2 \ge
\tT)\mathbin{?}(1-\distance{\code{q}}\code{[i]})\mathbin{:}0$'', where the
alignment actually depends on which branch is taken during the execution. 

To tackle the challenges, \tool generates an \emph{alignment template} for each
sampling instruction; a template is a numerical expression with ``holes''
whose values are to be searched for in later stages. For example, the
template generated for $\eta_2$ in GapSVT is 
\begin{align*}
(\code{q[i] + $\eta_2$ $\ge$ $\tT$})\mathbin{?} & (\hole[0] + \hole[1]
\times \distance{$\tT$} + \hole[2] \times
\distance{\code{q}}\code{[i]})\mathbin{:}\\
& (\hole[3] + \hole[4] \times \distance{$\tT$} + \hole[5] \times
\distance{\code{q}}\code{[i]})
\end{align*}
where $\hole[0]-\hole[5]$ are symbolic coefficients to be found later.

In general, for each sampling command $\eta = \lapm{\real}$, \tool first uses
static program analysis to find a set of relevant program expressions, denoted by
\exprset, and a set of relevant program variables, denoted by \varset (as
described shortly). Second, it generates an alignment template as follows:
$$ \alignment_{\exprset} ::=  
\begin{cases}
e_0 \ ?\ \alignment_{\exprset \setminus \{e_0\}}\mathbin{:}\alignment_{\exprset \setminus \{e_0\}} \text{, when } \exprset=\{e_0,\cdots\}\\
\hole_0 + \sum_{v_i\in \varset} \hole_i \times v_i \text{with fresh $\hole_0,\cdots,\hole_{|\varset|}$} \text{, otherwise }
\end{cases}
$$
where $\hole$ denotes coefficients (``holes'') to be filled out out by later
stages and each of them is generated fresh.

To find proper \exprset and \varset, our insight is that the alignments serve
to ``cancel out'' the differences between two related runs (i.e., to make all
assertions pass).  Algorithm~\ref{alg:generate_template} follows the insight to
compute  \exprset and \varset for each sampling instruction: it takes
$\Gamma_s$, the typing environment right before the sampling instruction and $A$,
all assertions in the transformed code, as inputs.  It also assumes an oracle
\code{Depends($e, x$)} which returns $\true$ whenever the expression $e$
depends on the variable $x$.  We note that the oracle can be implemented as
standard program dependency analysis~\cite{aho1986compilers,ferrante1987} or
information flow analysis~\cite{Bergeretti:1985:IDA:2363.2366}; hence, we omit
the details in this paper.
\begin{algorithm}[ht]
\setstretch{0.9}
\SetKwProg{Fn}{function}{\string:}{}
\SetKwFunction{Depends}{Depends}
\SetKw{Break}{break}
\SetKwFunction{GenerateTemplate}{GenerateTemplate}
\SetKwInOut{Input}{input}
\DontPrintSemicolon
\Input{ $\env_s$: typing environment at sampling command \\ 
$A$: set of the generated assertions in the program} 
\Fn{\GenerateTemplate{$\env_s$, $A$}}{
$\exprset \gets \emptyset$, $\varset \gets \emptyset$\;
\ForEach{$\assert{e} \in A$}{
    \If{$\Depends(e, \eta)$}{\label{line:template_depends_eta}
        \If{$\assert{e}$ is generated by \ruleref{T-If}}{
            $e' \gets $ the branch condition of \code{\textbf{if}}\;
                $\exprset \gets \exprset \cup \{e'\}$\;
            
        }
        \ForEach{$v \in Vars\cup \{e_1[e_2] | e_1[e_2] \in e\}$ }{
            \If{$\Gamma_s\not\proves v:\basety_0 \land \Depends(e, v)$}{
                $\varset \gets \varset \cup \{v\}$ \;
            }
        }
    }
}
\ForEach{$e \in \exprset\cup \varset$ }{
remove $e$ from \exprset and \varset if not in scope\;
}
\Return \exprset,\varset;
}
\caption{Template generation for $\eta := \lapm{r}$}
\label{alg:generate_template}
\end{algorithm}    

The algorithm first checks (at
line~\ref{line:template_depends_eta}) if aligning $\eta$ has a chance to make
an assertion pass. If so, it will increment $\exprset$ and $\varset$ as
follows. For \exprset, we notice that only for the assertions
generated by rule~\ruleref{T-If}, depending on the branch condition allows the
alignment to have different values under different branches. Hence, we add the
branch condition to $\exprset$ in this case. 
For \varset, our goal is to use the alignment to ``cancel'' the differences
caused by other variables and array elements such as $q[i]$ used in
$e$. Hence, we only need to consider \distance{$v$} if 
\begin{inparaenum}[(1)] 
\item $v$ is different between two related runs (i.e., $\Gamma_s\not\proves v:\basety_0$)
and 
\item $v$ contributes the assertion (i.e., $e$ depends on $v$).
\end{inparaenum}

Finally, the algorithm performs a ``scope check'': if any element in $\exprset$
or $\varset$ contains out-of-scope variables, then the element is excluded; for
example, $\eta_1$ should not depend on $q[i]$ in GapSVT since $q[i]$,
essentially an iterator of $q$, is not in scope at that point.

Consider $\eta_1$ and $\eta_2$ in GapSVT. The assertions in the translated programs are 
(we only list
the assertion in the $\true$ branch since the constraint in $\false$ branch is
symmetric)
:
\begin{enumerate}
    \item \code{\assert{q[i] + $\eta_2$ + \distance{q}[i] + \distance{$\eta_2$} $\geq$ $\tT$ + \distance{$\tT$}}}
    \item \code{\assert{\distance{q}[i] + \distance{$\eta_2$} - \distance{$\tT$} = 0}}
\end{enumerate}

For $\eta_1$, we have $\env_s = \{\code{q}: *\}$ (we omit the base types
and the variables that have $0$ distance for brevity) and both assertions
depend on $\eta_1$. Since both assertions depend on $\eta_1$ and \code{q[i]},
Algorithm~\ref{alg:generate_template} adds \code{\distance{q}[i]} into
$\varset$. Moreover, assertion (1) is generated by rule~\ruleref{T-If}.
Thus,the algorithm adds \code{q[i] + $\eta_2$ $\ge$ $\tT$} into $\exprset$.
Finally, since \code{q[i]} is out of scope at the sampling instruction, expression
using \code{q[i]} and variable \code{q[i]} are excluded, resulting \varset=\{ \} and
\exprset=\{ \}.

For $\eta_2$, we have $\env_s$ = \{\code{q}: \code{*}, $\tT$: \code{*}\}.  Since
both assertions depend on $\eta_2$ and \code{q[i]} and $T$,
Algorithm~\ref{alg:generate_template} adds \code{\distance{q}[i]} and
\distance{$\tT$} into $\varset$. Similar to $\eta_1$, the algorithm also adds
\code{q[i] + $\eta_2$ $\ge$ $\tT$} into $\exprset$.  Finally, all
expressions and variable are in scope, resulting
\varset=\{$\code{\distance{q}[i]}$, \distance{$\tT$}\} and \exprset=\{\code{q[i]
+ $\eta_2$ $\ge$ $\tT$} \}.

\subsection{Function Signature Rewrite}
Finally, \tool rewrites the function signature
to reflect the extra parameters and holes introduced in the transformed code. In general, $M(inp)$
is transformed to a new function signature $M'(inp,\distance{$inp$},sample,\hole)$ where $\distance{$inp$}$ are
the distance variables associated with inputs whose distance is not zero (e.g.,
$\distance{$q$}$ is associated with $q$ in GapSVT), $sample$ is a list of random
values used in $M$, and $\hole$ are the missing holes in alignment templates.

\subsection{Shadow Execution}\label{sec:shadow_execution}

To tackle challenging mechanisms such as Report Noisy Max~\cite{Dwork06diffpriv}, \tool uses \emph{shadow execution}~\cite{shadowdp}. Intuitively, the shadow execution tracks another program execution where the injected noises are always the same as those in the original execution. Therefore, values computed in the shadow execution incur no privacy cost. The aligned execution can then switch to shadow execution when certain conditions are met, allowing extra permissiveness~\cite{shadowdp}.

Supporting shadow execution only requires a few modifications: 
\begin{enumerate}
    \item Expressions will have a pair of distances ($\pair{\first{\dexpr}}{\second{\dexpr}}$), where the extra distance $\second{\dexpr}$ tracks the distance in the shadow execution; 
    \item Since the branches and loop conditions in shadow execution are not aligned, they might diverge from the original execution. Hence, a separate shadow branch/loop is generated to correctly update the shadow distances for the variables. %
\end{enumerate} 

Since the extended transformation rules largely follow the corresponding typing rules of ShadowDP, we present the complete set of rules with detailed explanations in the \appendixref. 

\move{One nontrivial task in transformation, though, is to extend the template generation algorithm (Algorithm~\ref{alg:generate_template}) to generate extra \emph{selectors} for each sampling instruction (e..g, $\eta := \lapm{\real}$). Intuitively, a selector expression $\select$ with the following syntax decides if the aligned or shadow execution is picked:
\[
\begin{array}{lccl}
\text{Var Versions} & k &\in &\{\alignd, \shadow\} \\
\text{Selectors} & \select &::=\; &e\ ?\ \select_1:\select_2 \mid k \\
\end{array}
\]

To automatically generate selectors, we extend Algorithm~\ref{alg:generate_template} to return a selector template $\select$. The definition is similar to the alignment template, where the value can depend on the branch conditions:
\[
\select_{\exprset} ::= 
\begin{cases}
e_0 \ ?\ \select_{\exprset \setminus \{e_0\}}\mathbin{:}\select_{\exprset \setminus \{e_0\}} \text{, when } \exprset=\{e_0,\cdots\}\\
\hole \text{ with fresh $\hole$} \quad\quad\ \ \ \ \text{, otherwise }
\end{cases}
\]
Compared with other holes ($\hole$) in the alignment template ($\alignment_{\exprset}$), the
only difference is that $\hole$ in $\select_{\exprset}$ has Boolean values representing whether to switch to shadow execution ($\shadow$) or not ($\alignd$). 
Hence, all holes in Alignment and Selector Temples are automatically generated in the same way (Section~\ref{sec:tool}).}

\subsection{Soundness}\label{sec:soundness}

\tool enforces a fundamental property: suppose $M(inp)$ is transformed to
$M'(inp,\distance{$inp$},sample,\hole)$, then $M(inp)$ is differentially private
if there is a list of values of $\hole$, such that all assertions in $M'$ hold
for all $inp,\distance{$inp$},\noise$.
Recall that an alignment template $\alignment$ is a function of $\hole$. Hence,
we have a concrete alignment $\alignment(\hole)$ (i.e., a proof) when such values of $\hole$ exist.

We build the soundness of \tool based on that of ShadowDP~\cite{shadowdp}. The main difference is that
ShadowDP requires every sampling command $\eta :=
\lapm~\real$ to be manually annotated. Thus, we can easily rewrite a program $M$ in \tool to a program
$\tilde{M}$ in ShadowDP by adding the following annotations:\\
\centerline{$\eta := \lapm~\real \rightarrow \eta :=
\lapm~\real;~\circ;~\alignment_\eta(\hole)$\quad \textsc{(\tool to ShadowDP)}}

where $\alignment_\eta$ is the alignment template for $\eta$.
\hl{We formalize the main soundness results next; the full proof can be found in the \extraref.}

\begin{theorem}[Soundness]\label{thm:soundness} Let $M$ be a mechanism written
in \tool.
With a list of concrete values of $\hole$, let $\tilde{M}$ be the corresponding mechanism in ShadowDP by rule
\ruleref{\tool to ShadowDP}. If (1) $M$ type checks, i.e., $\proves
\flowrule{\env}{M}{M'}{\env'}$ and (2) the assertions
in $M'$ hold for all inputs. Then $\tilde{M}$ type checks in ShadowDP, and the assertions in $\tilde{M}'$ (transformed from $\tilde{M}$ by ShadowDP) pass.
\end{theorem}

\begin{theorem}[Privacy]\label{thm:privacy}
With exactly the same notation and assumption as Theorem~\ref{thm:soundness}, $M$ satisfies $\epsilon$-differential privacy.
\end{theorem}

\section{Proof and Counterexample Generation}\label{sec:tool}

Recall that the transformed source code has the form of the following: $M'(inp, \distance{$inp$},
\noise, \hole)$.  For brevity, Let  $I$ denote a triple of $(inp,
\distance{$inp$}, \noise)$, and $C$ denote a counterexample in the form of
$C = (inp, inp', o)$ as defined in Section~\ref{sec:proof_and_counterexample}.
Proof/counterexample generation is divided into two
tasks:

\begin{itemize}
    \item Proof Generation: find an instantiation of $\hole$ such that assertions
in $M'$ never fail for any input $I$, or
    \item Counterexample Generation: find an instantiation of $I$, such that no
$\hole$ exists to make all assertions in $M'$ pass, and then construct a
counterexample $C$ based on $I$.
\end{itemize} 

\begin{figure}
\centering
\includegraphics[width=.9\columnwidth]{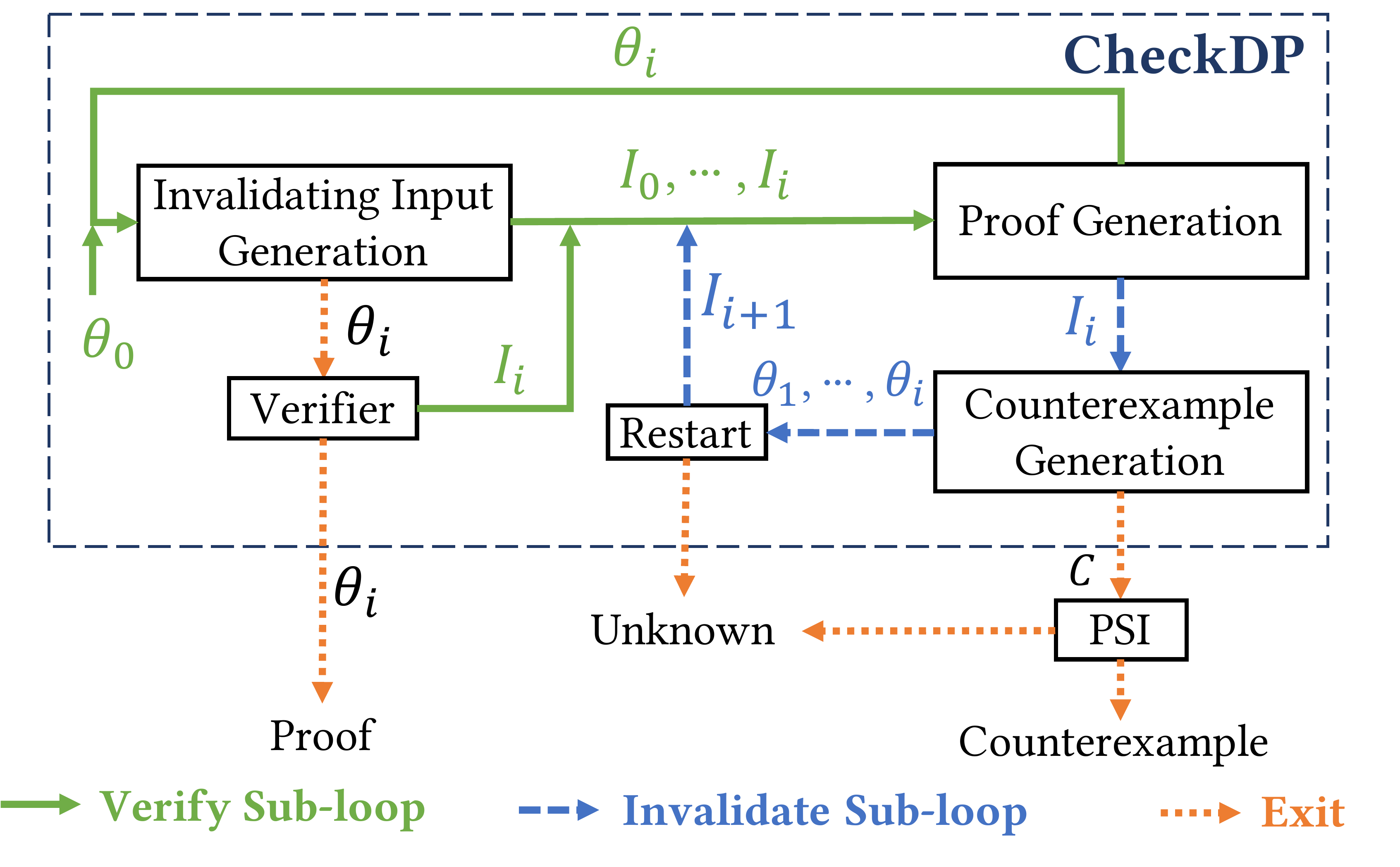}
\vspace{-3ex}
\caption{Overview of the  verify-invalidate loop of \tool.}\label{fig:tool_overview}
\vspace{-3ex}
\end{figure}

The key challenge here is the infinite search space of both $\hole$ and $I$.
Our insight is to use a verify-invalidate loop, as depicted in Figure~\ref{fig:tool_overview}, to improve $\hole$ and $I$ after each iteration. At a high-level, the iterative process involves two sub-loops: the (green) verify sub-loop generates proofs, and the (blue) invalidate sub-loop generates counterexamples. Moreover, the two sub-loops are integrated: starting from a default $\hole_0$ where
$\hole_0[i]=0.~\forall i$, the procedure generates sequences of proofs and
invalidating inputs in the form of $\hole_0,I_0,\hole_1,I_1,\cdots$. The final
$\hole_k$ or $I_k$ is used to construct proof or counterexamples
correspondingly.

\subsection{Verify Sub-Loop}
The verify sub-loop that involves Invalidating Input Generation and Proof Generation components is responsible of generating a sequence of improving alignments $\hole_0,\hole_1,\cdots,\hole_i$ such that, if the mechanism is correct, $\hole_i$ is a privacy proof (i.e, $\forall I.~M'(I, \hole_i)$). 

\paragraph{Invalidating Input Generation} This component takes a proof candidate $\hole_i$ and then tries to find an input $I_i$ such that
$
    \neg M'(I_i, \hole_i)
$
(meaning that at least one assertion in $M'$ fails).

Intuitively, $\hole_i$ is the currently ``best'' proof candidate (initially, a default
null proof $\hole_0 = [0,\cdots]$ is used to bootstrap the process) \hl{that is able to validate all previously found inputs ($I_0,\cdots, I_{i-1}$)}. An input $I_i$, if any, shows that $\hole_i$ is in fact not a valid proof (recall that a proof
needs to ensure $M'(I, \hole_i)~\forall I$). Hence, we call such $I_i$ an
\emph{invaliding input} of $\hole_i$ and feed it with all previously
identified invalidating inputs to the Proof Generation component following the
``Verify Sub-loop'' edge. 

\hl{Take GapSVT (Figure~\ref{alg:gapsvt}) for example. Since the initial null proof $\hole_0 = [0,\cdots]$ does not align any random variable, any input, say $I_0$, that diverges on the branch $q[i] + \eta_2 \ge T$ will trigger an assertion violation at Line~\ref{line:gapsvt_true_assertion}. Hence, the identified invalidating input $I_0$ is fed to the Proof Generation component.

}

\paragraph{Proof Generation}
This component takes in a series of invalidating inputs
$I_0, \cdots, I_i$ seen so far, and tries to find an proof candidate $\hole_i$ such
that:
$$
    M'(I_0, \hole_i) \land \cdots \land M'(I_i, \hole_i).
$$

Intuitively, the goal is to find a proof candidate $\hole_{i}$ that successfully ``covers'' all
invalidating inputs seen so far. Most likely, an improved proof candidate
$\hole_{i}$ \hl{that is able to align randomness for more inputs} is generated by the component. 
Then $\hole_{i}$ is fed back to the Invalidating Input Generation component, closing the loop. 

\hl{Consider the GapSVT example again. In order to align randomness for the invalidating input $I_0$, one possible $\hole_1$ is to align the random variable $\eta_2$ by $-\distance{$q$}[i]$ to cancel out the difference introduced by $q[i]$.
Note that this tentative proof $\theta_1$ does not work for \emph{all possible} inputs: it only serves as the ``best'' proof given $I_0$. With the Verify Sub-loop, such imperfect proof candidates enable the generation of more invalidating inputs, such as an invalidating input $I_1$ where the query answers are mostly below the threshold $T$ ($I_1$ invalidates $\hole_1$ since a privacy cost incurs whenever any branch is taken, which
eventually exhausts the given privacy budget). Therefore, a more general proof that leverages the conditional expression $q[i] + \eta_2 \ge T \mathbin{?} \bullet \mathbin{:} \bullet$ in the alignment template can be discovered by Proof Generation. For GapSVT, the Verify sub-loop eventually terminates with a correct proof (Section~\ref{sec:impexp}).

}

\paragraph{Exit Edges}
The verify loop has two exit edges. First, when no invalidating input is generated, $\hole_i$ is likely a valid proof.
Hence, $\hole_i$ is passed to a verifier with the following condition:
$
    \forall I.~M'(I, \hole_i).
$
Due to the soundness result (Theorem~\ref{thm:privacy}), we have a proof of
differential privacy when the verifier passes (the ``Exit'' edge from Verifier component). Otherwise,
\tool uses the counterexample returned by the verifier to construct $I_i$ (the
``Verify Sub-loop'' edge). We note that the verification step is required since KLEE, the symbolic executor that we use to find invalidating inputs, is unsound (i.e., it might miss an invalidating input) in theory; however, we did not experience any such unsound case of KLEE in our experience.

Second, the Proof Generation component might fail to find an alignment for $I_0, \cdots, I_i$, a case that will eventually occur for incorrect mechanisms. This exit edge leads to the invalidate sub-loop that we discuss next.

\subsection{Invalidate Sub-Loop}
\begin{figure}
\centering
\begin{subfigure}[b]{0.22\textwidth}
\captionsetup{width=\linewidth,format=hang}
\centering
\includegraphics[width=0.45\columnwidth]{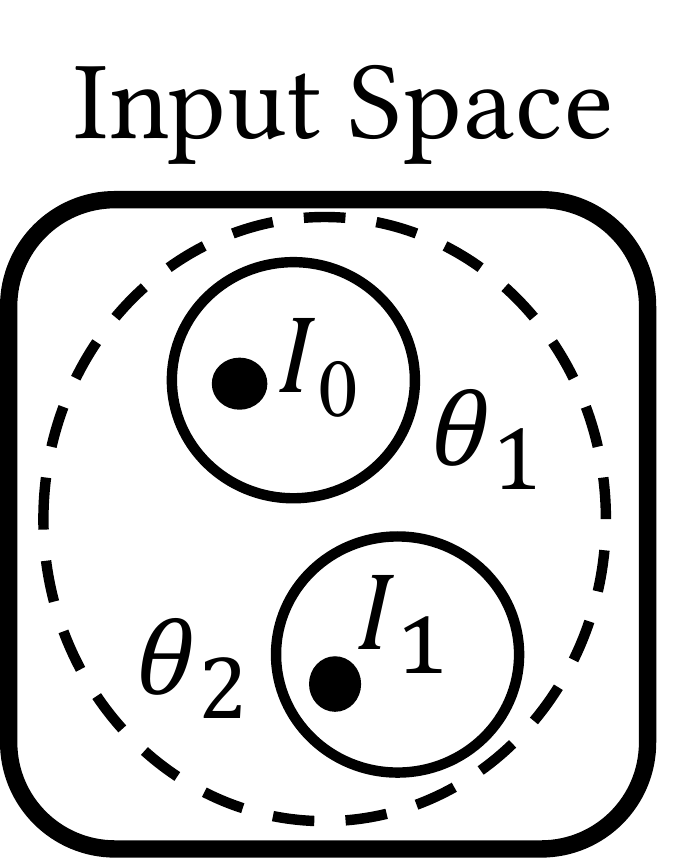}
\caption{A case where $\hole_i$ cannot be improved.}\label{fig:restart}
\end{subfigure}
\hspace{1em}
\begin{subfigure}[b]{0.22\textwidth}
\captionsetup{width=\linewidth,format=hang}
\centering
\includegraphics[width=0.45\columnwidth]{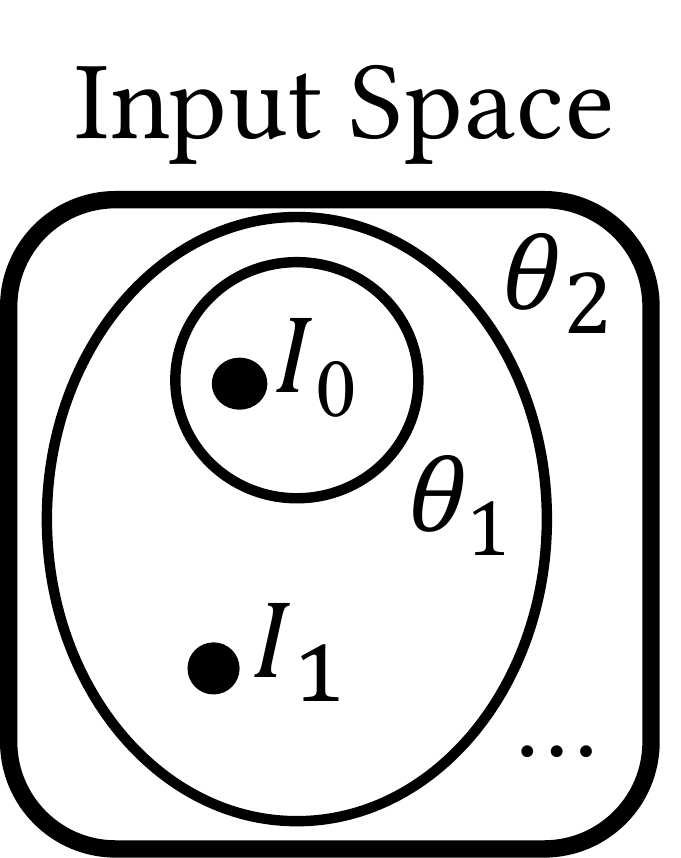}
\caption{Iteratively improving the alignment $\hole_i$.}\label{fig:improving_alignment}
\end{subfigure}
\vspace{-3ex}
\caption{Tentative alignments and invalidating inputs.}\label{fig:proof_generation}
\vspace{-3ex}
\end{figure}

The invalidate sub-loop involves Counterexample Generation and 
Restart;
it is responsible of generating \emph{one single} invalidating input $I$ such that, if the mechanism is incorrect, $I$ 
cannot be aligned (i.e, $\not\exists \theta.~M'(I,\theta)$). At first glance, it could be attempting 
to directly use $I_i$ from the Verify Sub-Loop. However, this is problematic both in theory and in practice:
 no alignment for $I_0, \cdots, I_i$ does not imply no alignment of $I_i$ alone. In practice, we found such a naive approach fails for BadSmartSum and BadGapSVT in Section~\ref{sec:impexp}.

\paragraph{Counterexample Generation} This component takes an invalidating input $I_i$ and then tries to find an alignment $\theta_i$ such that
$
    M'(I_i, \hole_i)
$
(meaning that $I_i$ is not a counterexample since it can be aligned by $\theta_i$).
For example, consider a corner case in Figure~\ref{fig:restart}, where Proof
Generation fails to find a common proof of both $I_0$ and $I_1$, but each of
$I_0$ and $I_1$ has a proof (illustrated by the two solid circles around them).
Mostly likely, this occurs when the program being analyzed is incorrect (hence,
no common proof) but neither $I_1$ nor $I_2$ is a good candidate for
counterexample of differential privacy, since each of them can be aligned in
isolation.

\paragraph{Restart}
This component is symmetric to the Invalidating Input Generation component in the verify sub-loop: it takes all previously found proof
candidates $\hole_1, \cdots, \hole_i$ and tries to find an invalidating input
$I_{i+1}$ such that:
$$
\neg M'(I_{i+1},\hole_1)\land \cdots \neg M'(I_{i+1},\hole_i).
$$

If found, $I_{i+1}$ will intuitively be out of scope of all found proofs and
serve as a ``better'' invalidating input. In theory, we can close the invalidate sub-loop by feeding $I_{i+1}$ back to Counterexample Generation. However, doing so will make proof and counterexample generation isolated tasks. Instead, we take an integrated approach, which we discuss shortly, where the verify and invalidate sub-loops communicate to generate proofs and counterexamples in a more efficient and simultaneous way.

\paragraph{Exit Edges}
If no $\hole$ is found to prove $I_i=(inp,\distance{inp},sample)$,
a counterexample 
$
C = (inp, inp + \distance{$inp$}, M'(inp, \distance{$inp$}, \noise, \hole_0))
$ can be formed and sent to an external exact probabilistic solver PSI~\cite{psisolver}
for validation.  
In theory, the Restart component might fail to find a new invalidating input given $\hole_1, \cdots, \hole_i$. However, this ``unknown'' state never showed up in our experience.

\subsection{Integrating Verify and Invalidate Sub-Loops}

We integrate the verify and invalidate sub-loops as follows: following the ``Invalidate Sub-loop'' edge of the Proof Generation component, the latest invalidating
input $I_i$ (i.e., the ``best'' invalidating input so far) is passed to the
Counterexample Generation component to start the invalidate sub-loop. Moreover, the newly generated invalidating input $I_i$ from the Restart component is fed back to the Proof Generation component to start the verify sub-loop. 

We note that by the design of the verify-invalidate loop, it alternatively runs
Invalidating Input Generation and Proof Generation components. By doing so, the
proof keeps improving while the invalidating inputs are getting closer to a
true counterexample (since the most recent one violates a ``better'' proof).
More intuitively, consider an invalidating input $I_0$ as a point in the entire
input space, illustrated in Figure~\ref{fig:improving_alignment}. A proof
candidate $\hole_1$ is able to prove the algorithm for a subset of inputs
including $I_0$ (indicated by the circle around $I_0$). The Invalidating Input
Generation component then tries to find another invalidating $I_1$ that
violates $\hole_1$ (falls outside of the $\hole_1$ circle). Next, the Proof
Generation component finds better proof candidate $\hole_2$ which proves
(``covers'') both $I_0$ and $I_1$.

We also note that it is crucial to consider all invalidating inputs so far
rather than the last input $I_i$ in the Proof Generation component: the efficiency of our approach
crucially relies on ``improving'' the proofs quantified by validating
more invalidating inputs. Without the improving proofs, the iterative procedure
might fail to terminate in case shown in Figure~\ref{fig:restart}: the
procedure might repeat $I_0,\hole_1,I_1,\hole_2,I_0,\hole_1,\cdots$. This
is confirmed in our empirical study.

\paragraph*{Unknown State}

Due to the soundness result (Theorem~\ref{thm:privacy}), the program being
analyzed is verified whenever \tool returns with a proof. Moreover, a validated 
counterexample by PSI disproves an incorrect mechanism. However, two reasons might
lead to the ``unknown'' state in the Figure~\ref{fig:tool_overview}: the generated 
counterexample is invalid or the Restart component fails
to find a new invalidating input. However, for all the correct and incorrect
examples we explored, the unknown state never showed up.

\section{Implementation and Evaluation}\label{sec:impexp}
We implemented \tool in Python\footnote{Publically available at \url{https://github.com/cmla-psu/checkdp}.}. \move{Details of the \emph{Program Transformation} and \emph{Proof and Counterexample Generation} phases are as follows:}
The \emph{Program Transformation} phase is implemented as a trans-compiler from CheckDP code (Figure~\ref{fig:syntax}) to C code. Following the transformation rules in Figure~\ref{fig:trans_rules}, the trans-compiler tracks the typing environment, gathers the needed constraints for the expressions, and more importantly, instruments corresponding statements when appropriate. Moreover, it adds a final assertion $\assert{\vpriv{\priv} \le \priv_b}$ before each $\outcmd{\!}$ command, where $\priv_b$ is the annotated privacy bound to be checked. Once all assertions are generated, the trans-compiler generates one alignment template for each sampling instruction as described in Algorithm~\ref{alg:generate_template}. For the \emph{{Proof and Counterexample Generation}} phase (i.e., verify-invalidate loop in Section~\ref{sec:tool}), we used an efficient symbolic executor
KLEE~\cite{klee} for most tasks.
Due to limited support of unbounded lists in KLEE, we fix the length of lists to be 5 in our evaluation. %
Also, to speed up the search, KLEE is configured to exit once
an assertion is hit. \hl{We note that the use of KLEE is to \emph{discover} alignments and counterexamples, where alignments are eventually verified by our sound Verifier component with arbitrary array length; counterexamples are confirmed by PSI. Moreover, \tool automatically extends the array length until either a verified proof or verified counterexample is produced.}

Finally, we deploy a 
verification tool CPAChecker~\cite{beyer2011cpachecker} for the Verifier component in \tool, which is capable of
automatically verifying C programs with given configuration
(\emph{predicateAnalysis} is used). Note that CPAChecker is able to generate
counterexamples for a failed verification. If the verification fails 
(which did not happen in our evaluation), 
\tool can feed the counterexample
back to the Proof and Counterexample Generation component.
\begin{table*}
\centering
\small{
\caption{Detected counterexamples for the incorrect algorithms and comparisons with other sampling-based counterexample detectors. $\#t$ stands for $\true$ and $\#f$ stands for $\false$.\label{tab:counterexamples}}
\vspace{-2ex}
\small
\resizebox{\linewidth}{!}{
\begin{tabular}{ccccccc||ccc}
\Xhline{1.5\arrayrulewidth}
\multicolumn{1}{c}{\textbf{Mechanism}} & \code{q} & \code{q}$'$  & \multicolumn{1}{c}{\textbf{Extra Args}}& \multicolumn{1}{c}{\textbf{Output}} & \multicolumn{1}{c}{\textbf{Iterations}} & \multicolumn{1}{c||}{\textbf{Time(s)}} & \textbf{StatDP~\cite{Ding2018CCS}} & \textbf{DP-Finder~\cite{Bichsel2018CCS}} & \textbf{DiPC~\cite{barthe2020}} \\
\Xhline{1\arrayrulewidth}
BadNoisyMax & $[0, 0, 0, 0, 0]$ & $[-1, 1, 1, 1, 1]$  & N/A & $0$ & 3 &  5.7 & 11.2 & 2561.5 & N/A  \\
BadSVT1 & $[0, 0, 0, 0, 1]$ & $[1, 1, 1, 1, 0]$  & $T$: 0, $N$: 1 & \makecell{$[\#f, \#f, \#f, \#f, \#t]$} & 4 & 3.2 & 4.9 & 3847.5 (Semi-Manual) & N/A \\
BadSVT2 & $[0, 0, 0, 0, 1]$ & $[1, 1, 1, 1, -1]$  & $T$: 0, $N$: 1 & \makecell{$[\#f, \#f, \#f, \#f, \#t]$} & 4 & 2.0 & 15.6 & 4126.1 (Semi-Manual) & N/A  \\
BadSVT3         & $[0, 0, 0, 0, 1]$ & $[1, 1, 1, 1, -1]$  & $T$: 0, $N$: 1 &\makecell{$[\#f, \#f, \#f, \#f, \#t]$} & 4 & 2.1 & 9.1 & 3476.2 (Semi-Manual)   & 269   \\ 
BadGapSVT & $[0, 0, 0, 0, 0]$ & $[1, 1, 1, 1, -1]$  & $T$: 0, $N$: 1 & \makecell{$[0, 0, 0, 0,1]$} & 4 &  5.7 & 10.6 & 11611.6 (Semi-Manual) & N/A \\
BadAdaptiveSVT & $[0, 0, 0, 0, 2]$ & $[1, 1, 1, 1, -1]$  & $T$: 0, $N$: 1 & \makecell{$[0, 0, 0, 0, 17]$} & 8 & 14.2 & 
Search Failed %
& 
Search Failed %
& N/A
\\
Imprecise SVT         & $[0, 0, 0, 0, 1]$ & $[1, 1, 1, 1, -1]$ & $T$: 0, $N$: 1 &\makecell{$[\#f, \#f, \#f, \#f, \#t]$} & 4 & 8.6 & Search Failed %
& 
Search Failed %
& N/A
\\ 
BadSmartSum & $[0, 0, 0, 0, 0]$ & $[0, 0, 0, 1, 0]$  & $T$: 3, $M$: 4  & \makecell{$[0,0,0,0,0]$} & 4 & 6.3 & 22.4 (Semi-Manual)  & Search Failed & N/A \\
BadPartialSum & $[0, 0, 0, 0, 0]$ & $[0, 0, 0, 0, 1]$  & N/A & \makecell{$0$} & 3 & 3.7 & 3.8 & 1128.5 & N/A\\
\Xhline{1.5\arrayrulewidth} 
\end{tabular}}
\vspace{-1ex}
}
\end{table*}

\newcommand\noisymaxbranch{\ensuremath{\Omega_{NM}}}
\newcommand\svtbranch{\ensuremath{\Omega_{SVT}}}
\newcommand\adaptivetopbranch{\ensuremath{\Omega_{Top}}}
\newcommand\adaptivemiddlebranch{\ensuremath{\Omega_{Middle}}}

\begin{table*}
\centering
\small{
\caption{Alignments found for the correct algorithms. $\Omega_*$ stands for the branch condition in each mechanism, where $\noisymaxbranch = q[i] + \eta > bq \lor i = 0$, $\svtbranch = q[i]+\eta_2\ge\tT$, $\adaptivetopbranch = q[i]+\eta_2 - \tT\ge \sigma$, $\adaptivemiddlebranch = q[i]+\eta_3 - \tT\ge 0$\label{tab:alignments}}
\vspace{-3ex}
\small
\resizebox{\linewidth}{!}{
\begin{tabular}{cccccc||ccc}
\Xhline{1.5\arrayrulewidth}
\multirow{2}{*}{\textbf{Mechanism}} & \multicolumn{3}{c}{\textbf{Alignment}}  & \multirow{2}{*}{\textbf{Iterations}} & \multirow{2}{*}{\textbf{Time (s)}} & \multirow{2}{*}{\textbf{ShadowDP~\cite{shadowdp}}} & \multirow{2}{*}{\textbf{Coupling~\cite{Aws:synthesis} }} & \multirow{2}{*}{\textbf{DiPC~\cite{barthe2020}} } \\
\cline{2-4} & $\mathbf{\eta_1}$
& $\mathbf{\eta_2}$ & $\mathbf{\eta_3} $ & & & \\
\Xhline{1\arrayrulewidth}
\rule{0pt}{1ex}
ReportNoisyMax & $\ternary{\noisymaxbranch}{1-\distance{$q$}[i]}{0}$  & N/A & N/A & 10 & 69.3 & Manual & 22 & 193 \\
PartialSum & $-\distance{$sum$}$ & N/A & N/A & 2 & 5.6 & Manual & 14 & N/A \\
SmartSum & $-\distance{$sum$}-\distance{$q$}[i]$ & $-\distance{$q$}[i]$  & N/A & 6 & 6.8 & Manual & 255 & N/A \\
SVT & 1 & $ \ternary{\svtbranch{}}{1-\distance{$q$}[i]}{0}$ & N/A & 4 & 6.2 & Manual & 580 & 825\\
Monotone SVT (Increase)  & 0 & $\ternary{\svtbranch}{1-\distance{$q$}[i]}{0}$ & N/A  & 8 & 18.4 & N/A & N/A & N/A \\
Monotone SVT (Decrease) & 0 & $\ternary{\svtbranch}{-\distance{$q$}[i]}{0}$ & N/A  & 8 & 20.5 & N/A & N/A & N/A \\
GapSVT & 1 & $\ternary{\svtbranch}{1-\distance{$q$}[i]}{0}$ & N/A & 6 & 13.5 & Manual & N/A & N/A\\
NumSVT & 1 & $\ternary{\svtbranch}{2}{0}$ & $-\distance{$q$}[i]$  & 4 & 8.8 & Manual & 5 & N/A  \\
AdaptiveSVT & 1 & $ \ternary{\adaptivetopbranch}{ 1 - \distance{$q$}[i]}{0}$ & $\ternary{\adaptivemiddlebranch}{1 - \distance{$q$}[i]}{0}$  & 10 & 25.6 & N/A & N/A & N/A \\
\Xhline{1.5\arrayrulewidth}
\end{tabular}}
\vspace{-2ex}
}
\end{table*}

\subsection{Case Studies}

Aside from GapSVT, we also evaluate \tool on the standard benchmark used in previous mechanism verifiers~\cite{Aws:synthesis, lightdp, shadowdp} and counterexample generators~\cite{Ding2018CCS,Bichsel2018CCS},\footnote{We note that like all tools designed 
for privacy mechanisms (e.g., \cite{Aws:synthesis, lightdp, shadowdp, Ding2018CCS,Bichsel2018CCS}), the benchmark do not include iterative programs that are built on those privacy mechanisms, such as k-means clustering, k-medians, since they are out of scope.}
including correct ones such as NumSVT, PartialSum, and SmartSum, as well
as the incorrect variants of SVT reported
in~\cite{ninghuisparse} and BadPartialSum. To show the power of \tool and expressiveness of our template generation algorithm, we also evaluate on a couple of correct/incorrect mechanisms that, to the best of our knowledge, have not been proved/disproved by existing verifiers and counterexample generators. This set of mechanisms include: Sparse Vector with monotonic queries~\cite{ninghuisparse}, AdaptiveSVT (called Adaptive Sparse Vector with Gap in~\cite{freegap}) as well as 
new incorrect variants of SVT, AdaptiveSVT and SmartSum. 
For all mechanisms we explore, \tool is able to: 
\begin{inparaenum}[(1)]
\item provide a proof if it satisfies differential privacy, or
\item provide a counterexample if it violates the claimed level of privacy.
\end{inparaenum} 
Neither false positives nor false negatives were observed.
In this section, we discuss the new cases; detailed explanations can be 
found in the \appendixref.

\paragraph{Sparse Vector with Monotonic Queries}
The queries in some usages of SVT are monotonic. In such cases, a $\lapm{2N/\epsilon}$ noise (instead of $\lapm{4N/\epsilon}$ in SVT) is sufficient for $\epsilon$-privacy~\cite{ninghuisparse}.

\paragraph*{AdaptiveSVT, BadAdaptiveSVT and BadSmartSum}
Ding et al.~\cite{freegap} recently proposed a new variant of SVT which adaptively allocates privacy budget, saving privacy cost when \emph{noisy} query answers are much larger than the noisy threshold. The difference from standard (correct) SVT is that it first draws a $\eta_2 := \lapm{8N/\epsilon}$ noise (instead of $\lapm{4N/\epsilon}$ in SVT) and checks if the gap between noisy query and noisy threshold $\tT$ is larger than a preset hyper-parameter $\sigma$ (\code{\textbf{if}} $q[i]$ + $\eta_2$ - $\tT$ $\ge$ $\sigma$). If the test succeeds, the gap is directly returned, hence costing only $\epsilon/(8N)$ (instead of $\epsilon/(4N)$) privacy budget. Otherwise, it draws $\eta_3 := \lapm{4N/\epsilon}$ and follows the same procedure as SVT. %
We also create an incorrect variant called BadAdaptiveSVT. 
It directly releases the noisy query answer instead of the gap after the first test. %
Sampling-based methods can have difficulty detecting the privacy leakage because the privacy-violating branch of the BadAdaptiveSVT code is not executed frequently.
We also create an incorrect variant of SmartSum by releasing a noise-less sum of queries in an infrequent branch.
Details of SmartSum and this variant can be found in the \appendixref.

\paragraph*{SVT with Wrong Privacy Claims (Imprecise SVT)}
We also study another interesting yet quite challenging violation of differential privacy: suppose a mechanism satisfies $1.1$-differential privacy but claims to be $1$-differentially private. This slight violation requires precise reasoning about the privacy cost and  poses challenges for prior sampling-based approaches. We thus evaluate a variant of SVT, referred to as Imprecise SVT, which is $\epsilon=1.1$-differentially private but with an incorrect claim of $\epsilon=1$ ($\textbf{check}(1)$ in the signature).

\subsection{Experiments}

We evaluate \tool on a $\text{Intel}^{\text{\textregistered}}$
$\text{Xeon}^{\text{\textregistered}}$ E5-2620 v4 CPU machine with 64
GB memory. To compare \tool with the state-of-the-art tools, \hl{we either directly run tools on the benchmark when they are publicly available (including ShadowDP~\cite{shadowdp}, StatDP~\cite{Ding2018CCS} and DP-Finder~\cite{Bichsel2018CCS}), or cite the reported results from the corresponding papers (including Coupling~\cite{Aws:synthesis} and DiPC~\cite{barthe2020})}.\footnote{Default settings are used in our evaluation: 100K/500K samples for event
selection/hypothesis testing components of StatDP; 50 iterations for sampling and optimization components of DP-Finder where each iteration collects 409,600 samples on average.} 
\hl{For the latter case, we note that the numbers are for reference only, due to different settings, including hardware, used in the experiments.}

\paragraph*{Counterexample Generation}
Table~\ref{tab:counterexamples} lists the
counterexamples (i.e., a pair of related inputs and a feasible output that
witness the violation of claimed level of privacy) automatically generated by \tool for the
incorrect algorithms. For all incorrect algorithms, \tool is able to provide a
counterexample (validated by PSI~\cite{psisolver}) in 15 seconds and 8 iterations.\footnote{We 
note that the counterexample of BadSmartSum is validated on a slightly modified algorithm
since PSI does not support modulo operation.}

Notably, both StatDP and DP-Finder fail to find the privacy violations in BadSmartSum and
BadAdaptiveSVT, as well as the violation of $\epsilon=1$-privacy in Imprecise SVT after hours of searching.\footnote{For StatDP, we use 1000X of the default number of samples to confirm the failure.} This is due to the
limitations of sampling-based approaches. In certain cases, we can help these sampling-based algorithms by \emph{manually} providing proper values for the extra arguments that some of the mechanisms require ($4^{th}$ column of Table \ref{tab:counterexamples}). This extra advantage (labeled
Semi-Manual in the table) sometimes allows the sampling-based methods to find counterexamples. %
We note that \tool, in contrast, generates all inputs automatically.

\paragraph*{Verification}
Table~\ref{tab:alignments} lists the automatically generated proofs (i.e.,
alignments) for each random variable in the correct algorithms. Due
to the soundness of \tool, all returned proofs are valid. We note that
correct algorithms on average take more iterations (and hence, time) to
verify; still all of them are verified  within 70 seconds. \move{Notably, \tool in fact provides a more precise alignment $q[i]+\eta_2\ge\tT \mathbin{?}
1-\distance{q}[i]\mathbin{:}0$ (same as the one in GapSVT) than the less
precise (though still correct) alignment $q[i]+\eta_2\ge\tT \mathbin{?}
2\mathbin{:}0$ manually generated in~\cite{lightdp}}. Report Noisy Max is the only example that uses shadow execution; the selector generated is $\select = \ternary{q[i] + \eta_2 \ge bq \lor i = 0}{\shadow}{\alignd}$, the same as the manually generated one in~\cite{shadowdp}.

\paragraph*{Performance}
We note that all examples finish within 10 iterations. We contribute the efficiency to the reduced search space of Algorithm~\ref{alg:generate_template} (e.g., the alignment template for GapSVT only contains 7 ``holes'') as well as our novel verify-invalidate loop that allows verification and counterexample 
generation components to communicate in meaningful ways. Compared with StatDP and DP-Finder, \tool is more efficient on the cases where 
they do find counterexamples.
Compared with static tools~\cite{Aws:synthesis, barthe2020}, we note that \tool is much faster on BadSVT3, SmartSum and SVT. 
In summary, \tool is mostly more efficient compared to counterexample detectors and automated provers.

\section{Related Work}\label{sec:relatedwork}
\hl{
\paragraph*{Proving and Disproving Differential Privacy.} Concurrent works~\cite{barthe2020, farina2020coupled} also target  both proving and disproving differential privacy. Barthe et al.~\cite{barthe2020} 
identify a non-trivial class of programs where checking differential privacy is decidable. Their work also supports approximate differential privacy. However, the decidable programs only allow finite inputs and outputs, while \tool
is applicable to a larger class of programs.
Moreover, \tool is more scalable, as observed in our evaluation. Farina~\cite{farina2020coupled} builds a relational symbolic execution framework, which when combined with probabilistic couplings, is able to prove differential privacy or generate failing traces for SVT and its two incorrect variants. However, it is unclear if the employed heuristic strategies work on other mechanisms, such as Report Noisy Max. Moreover, \tool is likely to be more scalable since their approach treats both program inputs and proofs in a symbolic way, whereas in the novel verify-invalidate loop of \tool, either program inputs or proofs are concrete.
}

\paragraph*{Formal Verification of Differential Privacy.}
\move{Differential privacy has been a rich target for program verification. Compared
with all existing verification work, \tool is \emph{the first} that provides useful
information (i.e., counterexamples) for incorrect mechanisms. Next, we compare
\tool with existing work purely from the verification perspective.}

\hl{From the verification perspective,} \tool is mostly related to  LightDP~\cite{lightdp} and
ShadowDP~\cite{shadowdp} --  all use randomness alignment.
The type system of \tool is directly inspired by that of
\cite{lightdp,shadowdp}. However, the most important difference is that 
\tool is \emph{the first} that automatically generates alignment-based proofs;
both LightDP and ShadowDP assume manually-provided proofs.
As discussed in Section~\ref{sec:translation}, \tool also simplifies the previous type systems and defers all privacy-related checks to later stages. Both changes are important for automatically generating proofs and counterexamples.

Besides alignment-based proofs, probabilistic couplings and liftings~\cite{BartheCCS16, Barthe16,Aws:synthesis} have also been used in language-based verification of differential privacy.
Most notably, Albarghouthi and Hsu~\cite{Aws:synthesis}
proposed the first automated tool capable of generating \emph{coupling proofs} for complex mechanisms. Coupling proofs are known to be more general than alignment-based proofs, while alignment-based proofs are more light-weight.
Since CheckDP and \cite{Aws:synthesis} are built on different proof techniques, the proof generation algorithm in~\cite{Aws:synthesis} is not directly applicable in our context. Moreover, \cite{Aws:synthesis} does not generate counterexamples and we do not see an obvious way to extend the Synthesize-Verify loop of~\cite{Aws:synthesis} to do so
\move{ (a) CEGIS is used as a component rather than the overall Synthesize-Verify loop; (b) unlike our approach where the generated invaliding inputs are ``improving'', it is unclear if the Synthesize-Verify loop of~\cite{Aws:synthesis} has the same property, and (c) the encoding of DP into low-level constraints makes it difficult to ``decode'' constraint solutions back to DP-counterexamples (i.e., a pair of inputs)}.

With verified privacy mechanisms, such as SVT and Report Noisy Max, we still need to verify that the larger program built on top of them is differentially private. An early line of work \cite{Barthe12, BartheICALP2013, Barthe14, Fuzz, DFuzz} uses (variations of) relational Hoare logic and linear indexed types to derive differential privacy guarantees. For example,
Fuzz~\cite{Fuzz} and its successor DFuzz\cite{DFuzz} combine linear indexed types and lightweight dependent types to allow rich sensitivity analysis and then use the composition theorem to prove overall system privacy. We note that \tool and those systems are largely orthogonal: those systems rely on trusted mechanisms (e.g., SVT and Report Noisy Max) without verifying them, while \tool is likely less scalable; they can be combined for sophisticated verification tasks.

\paragraph*{Counterexample Generation}
\move{Existing tools for proving/verifying that an algorithm satisfies differential privacy are not helpful when applied to incorrect algorithms. For this reason, there has been interest in generating counterexamples of differential privacy. }
Ding et al. \cite{Ding2018CCS} and Bichsel et al. \cite{Bichsel2018CCS} proposed
counterexample generators that rely on \emph{sampling} -- running an algorithm hundreds of thousands of times to estimate the output distribution of mechanisms (this information is then
used to find counterexamples). The strength of these methods is that they do not
rely on external solvers, and more importantly, they are not tied to (the limitation of) any
particular proof technique (e.g., randomness alignment and coupling). However,
sampling also make the counterexample detectors imprecise and more likely to fail in some cases, as confirmed in the evaluation.

\move{\paragraph*{CEGIS} The verify-invalidate loop of \tool is inspired by Guided Inductive Synthesis (CEGIS)~\cite{CEGIS}, a well-known technique in the context of program synthesis. 
However, due to the new context of generating alignment-based proofs, one key challenge is to refine the search space of candidate proofs (Algorithm~\ref{alg:generate_template}) so that the search space covers useful proofs while still allows efficient search. Moreover, the invalidate sub-loop as well as how we integrate the verify and invalidate sub-loops to efficiently search for both proofs and counterexamples simultaneously is novel. With the goal of counterexample generation, invalidate inputs are no longer part of the machinery for verification. }

\section{Conclusions and Future Work}\label{sec:conclusion}
We proposed \tool, an integrated tool based on static analysis for automatically
proving or disproving that a mechanism satisfies differential privacy.
Evaluation shows that \tool is able to provide proofs for a number of algorithms, as well as counterexamples for their incorrect variants within 2 to 70 seconds.  Moreover, all generated proofs and counterexamples are validated. 

\hl{For future work, \tool relies on the underlying randomness alignment technique; hence it is subject to its limitations, including lack of support for $(\epsilon, \delta)$-differential privacy and renyi differential privacy~\cite{M2017:Renyi}. We plan to extend the underlying proof technique for other variants of differential privacy.

Moreover, subtle mechanisms such as PrivTree~\cite{privtree} and private selection~\cite{privateselection}, where the costs of intermediate results are dependent on the data but the cost of sum is data-independent, is still out of reach for formal verification (including \tool). 

Finally, \tool is designed for DP mechanisms, rather than larger programs built on top of them. An interesting area of future work is integrating \tool with tools like DFuzz~\cite{DFuzz}, which are more efficient on programs built on top of DP mechanisms (but don't verify the mechanisms themselves).}

\ifack
\section*{Acknowledgments}
We thank the anonymous reviewers for their insightful feedbacks. This work was supported by NSF Awards CNS-1702760.
\fi
\balance
\bibliographystyle{ACM-Reference-Format}
\bibliography{diffpriv}

\ifappendix
\clearpage
\appendix
\section{\tool Semantics}
Let $A$ be a discrete set. The set of \emph{sub-distributions} over $A$, written $\dist(A)$, to be the set of functions $\mu: A\rightarrow [0,1]$ such that $\sum_{a\in A} \mu(a) \leq  1$. The reason to use sub-distributions instead of distributions (those $\mu$ such that  $\sum_{a\in A} \mu(a) =  1$) is that sub-distributions give rise to an elegant semantics for programs that do not necessarily terminate \cite{Kozen81}. 
We use $\dgdist_a$ to represent the degenerate distribution $\mu$ that $\mu(a)=1$ and $\mu(a')=0$ if $a'\not=a$.
Moreover, we define monadic functions $\unitop$ and $\bindop$ functions to
formalize the semantics for commands:
\begin{align*}
\unitop &: A \rightarrow \dist(A) \defn \lambda a.~\dgdist_a \\ 
\bindop &: \dist(A)\rightarrow(A \rightarrow \dist(B))\rightarrow\dist(B)  \\
        &\defn \lambda \mu.~\lambda f.~(\lambda b.~\sum_{a\in A} (f~a~b)\times \mu(a))
\end{align*}
That is, $\unitop$ takes an element in $A$ and returns the Dirac distribution
where all mass is assigned to $a$; $\bindop$ takes $\mu$, a distribution on
$A$, and $f$, a mapping from $A$ to distributions on $B$ (e.g., a conditional
distribution of $B$ given $A$), and returns the corresponding marginal
distribution on $B$. This monadic view avoids cluttered definitions and proofs
when probabilistic programs are involved.

\section{Shadow Execution}

We show how to extend the program transformation in Figure~\ref{fig:trans_rules} to support shadow execution. At a high level, the extension encodes the selectors (which requires manual annotations in ShadowDP~\cite{shadowdp}) and integrates them with the generated templates. With the extra ``holes'' in the templates, the verify-invalidate loop will automatically find alignments (including selectors)/counterexamples. The complete set of transformation rules with shadow execution is shown in Figure~\ref{fig:trans_rules_shadow}, where the extensions are highlighted in gray.

\paragraph*{Syntax and Expressions}
Since a new shadow execution is tracked, types for each variable would be expanded to include a pair of distances $\pair{\first{\dexpr}}{\second{\dexpr}}$. More specifically, the types should now be defined as: $\tau ::=\; \tyreal_{\pair{\first{\dexpr}}{\second{\dexpr}}} \mid \bool \mid \tylist~\tau$.

With the modified types, corresponding modifications to the transformation rules for expressions are straightforward and minimal: the handling of shadow distances are essentially the same as that of aligned distances.

\paragraph*{Normal Commands}
Following the type system of ShadowDP, a program counter $\pc$ $\in$ $\{ \top, \bot \}$ is introduced to each transformation rule for commands to capture potential divergence of shadow execution. Specifically, $\pc \proves c \transform c'$.  $\pc = \top$ (resp. $\bot$) means that the branch / loop command might diverge in the shadow execution (resp. must stay the same). The value of $\pc$ is used to guide how each rule should handle the shadow distances (e.g., \ruleref{T-Asgn}), which we will explain shortly. Therefore, another auxiliary function \code{updatePC} is added to track the value of \pc.

Compared with the type system of ShadowDP,
the first major difference is in \ruleref{T-Asgn}. If  $\pc = \bot$, shadow distances are handled as the aligned distances. However, when $\pc = \top$ (shadow execution diverges), it updates the shadow distance of the variable to make sure the value in shadow execution (i.e., $x + \second{\distance{$x$}}$) remains the same after the assignment. For example, Line~\ref{line:noisymax_bq_shadow} in Figure~\ref{alg:noisymax} is instrumented to maintain the value of \code{bq} in the shadow execution ($\code{bq} + \second{\distance{\code{bq}}}$), so that the branch at Line~\ref{line:noisymax_shadow_condition} is not affected by the new assignment of \code{bq}.

\newcommand\starexec[1]{\llparenthesis {#1} \rrparenthesis^\star}
\begin{figure}[ht]
\setstretch{0.6}
\raggedright
\small
\begin{mathpar}
\starexec{\real,\Gamma} = \real  
\quad
\starexec{\true,\Gamma} = \true
\quad
\starexec{\false,\Gamma} = \false
\and 
\starexec{x,\Gamma} = 
\begin{cases}
x+\second\nexpr &\text{, if } \Gamma \proves x : \tyreal_{\pair {\first\nexpr} {\second\nexpr}} \cr
x &\text{, else } 
\end{cases}
\and
\starexec{e_1~\op~e_2,\Gamma} = \starexec{e_1,\Gamma}~\op~\starexec{e_2,\Gamma}
\text{ where } \op = \oplus\cup \otimes \cup \odot
\\
\starexec{e_1[e_2],\Gamma} =~
\begin{cases}
e_1[e_2]+\second{\distance{$e_1$}}[e_2] &\text{, if } \second{\Gamma} \proves e_1 : \tylist~\tyreal_* \cr
e_1[e_2] &\text{, else} 
\end{cases}
\and 
\starexec{e_1::e_2,\Gamma} = 
\starexec{e_1,\Gamma}::\starexec{e_2,\Gamma} 
\and
\starexec{\neg e,\Gamma} = \neg \starexec{e, \Gamma}
\and
\starexec{e_1\mathbin{?}e_2:e_3,\Gamma} = \starexec{e_1}\mathbin{?}\starexec{e_2,\Gamma}:\starexec{e_3,\Gamma}
\\\\
\starexec{\skipcmd,\Gamma} = \skipcmd 
\and
\inferrule{
\starexec{c_1;\Gamma} = c_1'
\quad 
\starexec{c_2;\Gamma} = c_2'
}{
\starexec{c_1;c_2,\Gamma} = c_1';c_2'
}
\and
\starexec{x:=e,\Gamma} = ({\second{\distance{$x$}}} := \starexec{e,\Gamma}-x)
\and
\inferrule{
\starexec{c_i, \Gamma} = c_i' \quad i \in \{1,2\}
}
{
\starexec{\ifcmd{e}{c_1}{c_2},\Gamma} = \ifcmd{\starexec{e,\Gamma}}{c_1'}{c_2'}}
\and
\inferrule{
\starexec{c,\Gamma}= c'
}{\starexec{\whilecmd{e}{c},\Gamma} = \whilecmd{\starexec{e,\Gamma}}{c'}}
\end{mathpar}
\caption{Transformation of expressions and commands for aligned and shadow execution, where $\star \in \{\alignd, \shadow\}$.}
\label{fig:shadowrules}
\end{figure}

As previously explained, a separate shadow branch / loop has to be generated to correctly track the shadow distances of the variables. More specifically, Rules~\ruleref{T-If} and~\ruleref{T-While} is extended to include an extra shadow execution command $\second{c}$ \emph{when \pc\ transits from $\bot$ to $\top$}. The shadow execution is constructed by an auxiliary function $\shadowexec{c,\env}$, as defined in Figure~\ref{fig:shadowrules}, which is the same as the ones in ShadowDP~\cite{shadowdp}. It essentially replaces each variable with its correspondence (e.g., variable $x$ to $x + \second{\distance{$x$}}$), as is standard in self-composition~\cite{barthe2004, terauchi2005}. Note that the value of an expression $e$ in an aligned execution (i.e., $\aligndexec{e,\env}$ used in Rules~\ruleref{T-If} and~\ruleref{T-While}) are defined in a similar way.

\paragraph*{Sampling Commands}

The most interesting rule is \ruleref{T-Laplace}. In order to enable the automatic discovery of the selectors, our \code{GenerateTemplate} algorithm needs to be extended to return a selector template $\select$. Intuitively, a selector expression $\select$ with the following syntax decides if the aligned or shadow execution is picked:
\[
\begin{array}{lccl}
\text{Var Versions} & k &\in &\{\alignd, \shadow\} \\
\text{Selectors} & \select &::=\; &e\ ?\ \select_1:\select_2 \mid k \\
\end{array}
\] 

The definition of the selector template is then similar to the alignment template, where the value can depend on the branch conditions:
\[
\select_{\exprset} ::= 
\begin{cases}
e_0 \ ?\ \select_{\exprset \setminus \{e_0\}}\mathbin{:}\select_{\exprset \setminus \{e_0\}} \text{, when } \exprset=\{e_0,\cdots\}\\
\hole \text{ with fresh $\hole$} \text{, otherwise }
\end{cases}
\]

\clearpage
\newcommand\alignexec[1]{\llparenthesis {#1} \rrparenthesis^\alignd}
\begin{figure*}[!ht]
\small
\raggedright

\framebox{\textbf{Transformation rules for expressions with form $\env \proves e:\basety_{\pair{\first{\nexpr}}{\second{\nexpr}}}$}}
{\setstretch{0.7}
\begin{mathpar}
\inferrule*[right=(T-Num)]{ }
{ \env \proves \exprrule{\real}{\tyreal_{\pair{0}{\diff{\scriptstyle 0}} }}{\real}{\noconstraints}}
\and
\inferrule*[right=(T-Boolean)]{ }
{\env \proves \exprrule{b}{\bool}{b}{\noconstraints}}
\and
\inferrule*[right=(T-Neg)]
{\env\proves \exprrule{e}{\bool}{e'}{\constraints}}
{ \env \proves \exprrule{\neg e}{\bool}{e'}{\constraints}}
\and
\inferrule*[right=(T-Var)]{ \env(x) = \basety_{\pair{\first{\dexpr}}{\diff{\scriptstyle \second{\dexpr}}}} 
\quad
\third{\nexpr} = 
\begin{cases}
\third{\distance{$x$}} & \text{if } \third{\dexpr} = * \cr
0 & \text{otherwise}
\end{cases}
\quad
\star\in\set{\circ, \dagger}
}{\env \proves \exprrule{x}{\basety_{\pair{\first{\nexpr}}{\diff{\scriptstyle \second{\nexpr}}}}}{}{\true} }
\and
\inferrule*[right=(T-OPlus)]
{\env\proves \exprrule{e_1}{\tyreal_{\pair {\nexpr_1} {\diff{\nexpr_2}}}}{e_1'}{\constraints_1} \quad \env\proves \exprrule{e_2}{\tyreal_{\pair {\nexpr_3} {\diff{\scriptstyle \nexpr_4}}}}{e_2'}{\constraints_2}{}}
{\env \proves \exprrule{e_1 \oplus e_2}{\tyreal_{\pair {\nexpr_1 \oplus \nexpr_3} {\diff{\scriptstyle \nexpr_2 \oplus \nexpr_4 }}}}{e_1' \oplus e_2'}{\constraints_1 \land \constraints_2}}
\quad
\inferrule*[right=(T-OTimes)]
{\env\proves \exprrule{e_1}{\tyreal_{\pair {\nexpr_1} {\diff \scriptstyle{\nexpr_2}}}}{}{\constraints_1} \quad \env\proves \exprrule{e_2}{\tyreal_{\pair {\nexpr_3} {\diff{\scriptstyle \nexpr_4}}}}{}{\constraints_2}}
{\env \proves \exprrule{e_1 \otimes e_2}{\tyreal_{\pair 0 0}}{}{\constraints_1 \land \constraints_2 \land \inferrule{}{(\nexpr_1 = \nexpr_2 = \\\\ \diff{\nexpr_3 = \nexpr_4 =\ } 0)}}}
\and
\inferrule*[right=(T-ODot)]
{\env\proves \exprrule{e_1}{\tyreal_{\pair {\nexpr_1} {\diff{\scriptstyle \nexpr_2}}}}{}{\constraints_1} \quad \env\proves \exprrule{e_2}{\tyreal_{\pair {\nexpr_3} {\diff{\scriptstyle \nexpr_4}}}}{}{\constraints_2}}
{\env \proves \exprrule{e_1 \odot e_2}{\bool}{}{\constraints_1 \land \constraints_2 \land \inferrule{}{(e_1 \odot e_2) \Leftrightarrow (e_1 + \nexpr_1) \odot (e_2 + \nexpr_3) \diff{\land (e_1 \odot e_2) \Leftrightarrow (e_1 + \nexpr_2) \odot (e_2 + \nexpr_4)}} }}
\and
\inferrule*[right=(T-Cons)]
{\env\proves \exprrule{e_1}{\basety_{\pair {\nexpr_1} {\diff{\scriptstyle \nexpr_2}}}}{}{\constraints_1} \quad \env\proves \exprrule{e_2}{\tylist~\basety_{\pair {\nexpr_3} {\diff{\scriptstyle \nexpr_4}}}}{}{\constraints_2}}
{\env \proves \exprrule{e_1::e_2}{\tylist~\basety_{\pair {\scriptstyle \nexpr_3}{\diff{\scriptstyle \nexpr_4}}}}{}{\constraints_1 \land \constraints_2 \land \inferrule{}{(\nexpr_1 = \nexpr_2 = \diff{\nexpr_3 = \nexpr_4 =\ } 0})}}
\quad
\inferrule*[right=(T-Index)]
{\env\proves \exprrule{e_1}{\tylist~\tau}{}{\constraints_1}\quad \env\proves \exprrule{e_2}{\tyreal_{\pair{\nexpr_1}{\diff{\scriptstyle \nexpr_2}}}}{}{\constraints_2}}
{\env \proves \exprrule{e_1[e_2]}{\tau}{}{\constraints_1 \land \constraints_2 \land (\nexpr_1\ \diff{ = \nexpr_2} = 0)}}
\and
\inferrule*[right=(T-Select)]
{\env\proves \exprrule{e_1}{\bool}{}{\constraints_1} \quad \env\proves \exprrule{e_2}{\basety_{\pair {\nexpr_1} {\diff{\nexpr_2}}}}{}{\constraints_2} \quad \env\proves \exprrule{e_3}{\basety_{\pair {\nexpr_3} {\diff{\nexpr_4}}}}{}{\constraints_3}}
{\env \proves \exprrule{e_1\mathbin{?}e_2\mathbin{:}e_3}{\basety_{\pair {\nexpr_1}{\diff{\nexpr_2}}}}{}{\constraints_1 \land \constraints_2 \land \constraints_3\land (\nexpr_1=\nexpr_2\diff{=\nexpr_3=\nexpr_4})}}
\end{mathpar}
}
\framebox{\textbf{Transformation rules for commands with form $\pc \proves \flowrule{\env}{c}{c'}{\env'}$}}

\begin{mathpar}
\setstretch{0.9}
\inferrule*[right=(T-Asgn)]{ 
\env\proves\exprrule{e}{\basety_{\pair {\first{\nexpr}}{\diff{\scriptstyle \second{\nexpr}}}}}{}{\constraints} \quad
\configtwo{\first{\dexpr}}{\first{c}} = \begin{cases}
\configtwo{0}{\skipcmd} , \quad\ \text{ if } \first{\nexpr} == 0, \cr
\configtwo{*}{\first{\distance{$x$}} := \first{\nexpr}} , \text{ otherwise} 
\end{cases} \diff{
\configthree {\second{\dexpr}} {\second{c}} {c'} =
\begin{cases}
\configthree {0} {\skipcmd} {\skipcmd}, \quad\ \ \text{ if } \pc=\bot \land \second{\nexpr} = 0  \cr
\configthree{*}{\second{\distance{$x$}} := \second{\nexpr}}{\skipcmd}, \ \text{ if } \pc=\bot \land \second{\nexpr} \neq 0 \cr
\configthree {*} 
{\skipcmd} {\second{\distance{$x$}} := x + \second{\nexpr} - e}, \ \text{ otherwise }
\end{cases}
}}
{\diff{\pc}\proves \flowrule{\env}{x := e}{\assert{\constraints}; \diff{c'}; x := e; \first{c};\diff{\second{c}}}{ \env[x \mapsto \basety_{\pair {\first{\dexpr}} {\diff{\scriptstyle \second{\dexpr}}} }] }}
\vspace{-0.2cm}
\and
\inferrule*[right=(T-Seq)]
{\diff{\pc}\proves \flowrule{\env}{c_1}{c_1'}{\env_1} \quad \diff{\pc} \proves \flowrule{\env_1}{c_2}{c_2'}{\env_2}}
{\diff{\pc}\proves\flowrule{\env}{c_1;c_2}{c_1';c_2'}{\env_2}}
\and
\inferrule*[right=(T-Skip)]{ }
{\diff{\pc}\proves \flowrule{\env}{\skipcmd}{\skipcmd}{\env}}
\and
\vspace{-0.2cm}
\inferrule*[right=(T-Return)]
{ \env \proves \exprrule{e}{\basety_{\pair{\first{\nexpr}}{\diff{\scriptstyle \second{\nexpr}}}}}{}{\constraints}}
{\diff{\pc} \proves \flowrule{\env}{\outcmd{e}}{\assert{\constraints\land \first{\nexpr} = 0};\outcmd{e}}{\env}}
\and
\inferrule*[right=(T-While)]
{\inferrule{}{\diff{\pc}\proves\flowrule{\env \join \env_f}{c}{c'}{\env_f} \\\\ \diff{\pc' = \code{updatePC}(\pc, \env, e)} } \quad \inferrule{}{  \env, \env \join \env_f, \diff{\pc'} \Rrightarrow c_s \\\\ \env_f, \env \join \env_f, \diff{\pc} \Rrightarrow c''} \quad 
\diff{
\second{c} =
\begin{cases}
\skipcmd , \quad\text{ if } (\pc=\top \lor \pc'=\bot) \cr
\shadowexec{\whilecmd{e}{c},\Gamma\join \Gamma_f}, \quad\text{ else}
\end{cases}
}}
{\diff{\pc}\proves \flowrule{\env}{\whilecmd{e}{c}}{c_s; (\whilecmd{e}{(\assert{\alignexec{e, \env}};c';c''));\diff{\second{c}}}}{\env \join \env_f}}
\and
\inferrule*[right=(T-If)]
{
\diff{\pc}\proves \flowrule{\env}{c_i}{c_i'}{\env_i}  \quad \inferrule{}{\diff{\pc' = \code{updatePC}(\pc, \env, e)} \\\\ \env_i, \env_1 \join \env_2, \diff{\pc'} \Rrightarrow c''_i \quad i \in \{1, 2\}}  \quad \diff{{\second{c}} \!=\!
\begin{cases}
\skipcmd , \quad \text{ if } (\pc=\top \lor \pc'=\bot) \cr
{\shadowexec{\ifcmd{e}{c_1}{c_2},\Gamma_1\join\Gamma_2}},\text{ else}
\end{cases}}
}
{\diff{\pc}\proves \flowrule{\env}{\ifcmd{e}{c_1}{c_2}}{(\ifcmd{e}{(\assert{\alignexec{e,\env}};c_1';c_1'')}{(\assert{\lnot \alignexec{e,\env}};c_2';c_2'')});\diff{\second{c}}}{\env_1 \join \env_2}}
\and
\inferrule*[right=(T-Laplace)]
{\pc = \bot \qquad
\inferrule{}{\alignment, \diff{\select}=\code{GenerateTemplate}(\Gamma,\text{All Assertions}) \\\\ c_a = \assert{(\subst{(\eta + \alignment)}{\eta}{\eta_1} = \subst{(\eta + \alignment)}{\eta}{\eta_2} \Rightarrow \eta_1=\eta_2)} \\\\ \diff{\env' = \lambda x.~\pair{ \first{\dexpr} \join \second{\dexpr}}{\second{\dexpr}} \text{ where }\env(x)= \tyreal_{\pair{\first{\dexpr}}{\second{\dexpr}}}} } \quad \diff{\inferrule{}{ %
 c'= \{\first{\distance{$x$}} := 0 \mid  \env'(x)=\tyreal_{\pair{*}{\second{\dexpr}}} \land \env(x)=\tyreal_{\pair{0}{\second{\dexpr}}}\} \\\\
 c''=\{\first{\distance{$x$}} := \second{\nexpr} \mid \env' \proves x : \tyreal_{\pair{-}{\second{\nexpr}}}\}
 \\\\
 c_{\dexpr} = (\ifcmd{\select}{c'}{c''})}}}
{\diff{\pc} \proves \flowrule{\env}{\eta := \lapm{\real}}
{c_a; \ \eta := \noise[idx]; idx:= idx+1; \vpriv{\priv} := \diff{(\ternary{\select}{\vpriv{\priv}}{0})}+|\alignment|/r;\distance{$\eta$} := \alignment;\diff{c_{\dexpr}}}
{\env'[\eta \mapsto \tyreal_{\pair {*}{\diff{\scriptstyle 0}}}]}}
\end{mathpar}
\framebox{\textbf{Transformation rules for merging environments}}
\begin{mathpar}
\inferrule*
{\env_1 \sqsubseteq \env_2 \quad 
\inferrule*{}
{
\inferrule{}{\first{c}=\{\first{\distance{$x$}}:=0 \mid \Gamma_1(x)=\tyreal_{\pair {0} {\dexpr_1}} \land \Gamma_2(x)=\tyreal_{\pair * {\dexpr_2}}\} \\\\
\diff{\second{c}=\{\second{\distance{$x$}}:=0 \mid \Gamma_1(x)=\tyreal_{\pair {\dexpr_1} {0} } \land \Gamma_2(x)=\tyreal_{\pair {\dexpr_2} *}\} }} \quad \diff{c'=
\begin{cases}
\first{c};\second{c} & \text{ if } \pc=\bot \cr
\first{c} & \text{ if } \pc=\top \cr
\end{cases}}
}
}
{\env_1, \env_2, \diff{\pc} \Rrightarrow \diff{c'}}
\end{mathpar}
\vspace{-2ex}
\framebox{\textbf{PC update function}}
\vspace{-0.5em}
\begin{mathpar}
\diff{
\code{updatePC}(\pc, \Gamma, e)=
\begin{cases}
\bot \text{ , if } \pc=\bot\land \env\proves e:\tyreal_{\pair{-}{0}} \\
\top \text{ , else}
\end{cases}
}
\end{mathpar}
\vspace{-12px}
\caption{Rules for transforming probabilistic programs into deterministic ones with shadow execution extension. Differences that shadow execution introduce are marked in gray boxes.\label{fig:trans_rules_shadow}}
\end{figure*}

\clearpage

Compared with other holes ($\hole$) in the alignment template ($\alignment_{\exprset}$), the
only difference is that $\hole$ in $\select_{\exprset}$ has Boolean values representing whether to stay on aligned execution ($\alignd$), or switch to shadow execution ($\shadow$).

To embed shadow execution into \tool, the type system dynamically instruments an auxiliary command ($c_{\dexpr}$) according to the selector template $\select$. Once a switch is made ($\select = \shadow$), the distances of all variables are replaced with their shadow versions by this command. Moreover, the privacy cost $\vpriv{\priv}$ will also be properly reset according to the selector.

\section{Soundness Proof}
\tool's alignment-based proof system is built on that of ShadowDP~\cite{shadowdp}. At a high level, \tool automatically infers a proof in the form of alignment templates, so that the proof will be type-checked in a ShadowDP-like type system. Hence, given an inferred proof (i.e., concrete values of holes $\hole$ used in $\set{\alignment_\eta\mid \eta \in \rvars}$ or $\set{\alignment_\eta, \select_\eta \mid \eta \in \rvars}$ (with shadow execution), we can transform a program $M$ in  \tool to a program $\tilde{M}$ in ShadowDP according to the following rule:
\begin{mathpar}
\inferrule*[right=(\tool to ShadowDP)]{}{\eta := \lapm~\real \rightarrow \eta := \lapm~\real;~\select_\eta(\hole);~\alignment_\eta(\hole)}
\end{mathpar}

Without losing generality, we will proceed with the case with shadow execution (i.e., the type system $\Gamma$ tracks a pair of distances for both aligned and shadow executions), since a proof without shadow execution is subsumed by the one with shadow execution and a selector that always selects the aligned distances.

\textbf{Proof of Theorem~\ref{thm:soundness}}\\
Let $M$ be a mechanism written in \tool.
With a list of concrete values of $\hole$, let $\tilde{M}$ be the corresponding mechanism in ShadowDP by rule
\ruleref{\tool to ShadowDP}. If (1) $M$ type checks, i.e., $\proves
\flowrule{\env}{M}{M'}{\env'}$ and (2) the assertions
in $M'$ hold for all inputs. Then 
\begin{enumerate}
\item $\tilde{M}$ type checks in ShadowDP, and 
\item the assertions in $\tilde{M}'$ (transformed from $\tilde{M}$ by ShadowDP) pass.
\end{enumerate}

\begin{proof}

The proof is mostly straightforward due to the similarity between the type systems of \tool and ShadowDP. As stated in Section~\ref{sec:translation}, the only difference that requires extra work in the proof is that \tool only tracks if a variable has the same value in two related runs (with distance 0) or not (with distance $*$), while ShadowDP also allows distance of an arbitrary expression. To gap the potential difference, we define that $\Gamma'$ and $\Tilde{\Gamma}'$ are \emph{consistent} if 
\[\forall x\in \nvars \cup \rvars.~\aligndexec{x,\Gamma'}=\aligndexec{x,\Tilde{\Gamma}'}\land \shadowexec{x,\Gamma'}=\shadowexec{x,\Tilde{\Gamma}'}\]
Note that since we only need to convert \tool types to the (more expressive) ShadowDP types, such restriction of \tool types does not cause any issue.

First we show that if an expression $e$ of $M$ type checks with $\env$ in \tool, and all of the generated constraints $\constraints$ hold, then $e$ type checks with $\tilde{\env}$ in ShadowDP with an equivalent type (including distances), as long as $\env$ is consistent with $\Tilde{\env}$. 
We list a few interesting cases here. The proofs for other types of expressions are omitted since their rules in \tool are identical other than collecting static checks in ShadowDP as constraints. 
\begin{itemize}[leftmargin=5mm]
\item $e=x$: the interesting case is when $\Gamma(x)=\basety_{\pair * *}$ and $\Tilde{\Gamma}(x)=\basety_{\pair {\nexpr_1} {\nexpr_2}}$. We have the derived types are equivalent under $\Gamma$ and $\Tilde{\Gamma}$ by the consistency assumption.

\item $e= e_1?e_2:e_3$. Let $e_2, e_3$ be such that $\env \proves e_2: \tyreal_{\pair{\nexpr_1}{\nexpr_2}}$, $\env \proves e_3: \tyreal_{\pair{\nexpr_3}{\nexpr_4}}$. The T-Select rule restricts that $\nexpr_1=\nexpr_2=\nexpr_3=\nexpr_4$, which entails the requirement that $e_2$ and $e_3$ have the same type in the corresponding rule of ShadowDP.

\end{itemize}
Next, we show that if $\Gamma$ is consistent with $\Tilde{\Gamma}$ and $\proves \flowrule{\env}{M}{M'}{\env'}$, then $\proves \Tilde{\Gamma}~\{\tilde{M} \transform \tilde{M}'\}~\Tilde{\Gamma}'$ and $\Gamma'$ and $\Tilde{\Gamma}'$ are consistent. We proceed by rule induction on commands. For most rules, all assumptions in ShadowDP rules are guaranteed by the corresponding assertions in \tool, making them trivial cases. 
Next, we present the interesting cases and omit the rest ones.

\begin{itemize}
    \item $x := e$: let $\env \proves e: \tyreal_{\pair{\first{\nexpr}}{\second{\nexpr}}}$. The interesting case is when $\pc=\bot\land \second{\nexpr}\neq 0$. In \tool, since $x := e$ type checks in \tool, we know that $\Gamma'(x)=*$ and $\second{\distance{x}}$ is updated to $\second{\nexpr}$ after the transformed assignment. In ShadwoDP, we have $\Gamma'(x)=\second{\nexpr}$. Hence, $\Gamma'$ and $\Tilde{\Gamma}'$ are still consistent: $\shadowexec{x,\Gamma'}=x+\second{\nexpr}=x+\Tilde{\Gamma}'(x)=\shadowexec{x,\Tilde{\Gamma}'}$.
    
    \item $\eta  := g$: the assertion $c_a$ ensures that the corresponding static check succeeds in rule T-Laplace of ShadowDP. One notable difference  between \tool and ShadowDP is that since selector $\select$ is unknown statically, a branch $c_\dexpr$ is inserted to update the alignment of aligned execution. For consistency, checking $\shadowexec{x,\Gamma'}=\shadowexec{x,\Tilde{\Gamma}'}$ is trivial since 
    the shadow distances are updated in the same way as in ShadowDP.
    When $\select=\alignd$, the interesting case is when the distance of $x$ is promoted to $*$ (i.e., $\env'(x)=\tyreal_{\pair{*}{\second{\nexpr}}} \land \env(x)=\tyreal_{\pair{0}{\second{\nexpr}}} $. In this case, due to the inserted commands $c'$, $\aligndexec{x,\Gamma'}=x+\first{\distance{x}}=x=\aligndexec{x,\Tilde{\Gamma}'}$. When $\select=\shadow$, due to the inserted commands $c''$, $\shadowexec{x,\Gamma'}=x+\second{\distance{x}}=x+\second{\nexpr}=\aligndexec{x,\Tilde{\Gamma}'}$ where $\Gamma \proves x : \tyreal_{\pair{\_}{_,\second{\nexpr}}}$.
    Finally, the typing environment changes to $\Gamma'[\eta\mapsto \tyreal_{\pair {\nexpr_{\eta}} 0}]$ in ShadowDP, but since all nonzero distances are dynamically tracked in \tool, this becomes $\Gamma'[\eta\mapsto \tyreal_{\pair {\ast} 0}]$, which is the one given by CheckDP rule.
\end{itemize}

\end{proof}

\textbf{Proof of Theorem~\ref{thm:privacy}}\\
With exactly the same notation and assumption as Theorem~\ref{thm:soundness}, $M$ satisfies $\epsilon$-differential privacy.
\begin{proof}
This follows directly from Theorem~\ref{thm:soundness} and the soundness of ShadowDP (\cite{shadowdp}, Theorem 2) and the fact that $M$ and $\tilde{M}$ are semantically the same. 
\end{proof}

\begin{figure}[H]
\small 
\setstretch{0.9}
\raggedright
\noindent\rule{\linewidth}{2\arrayrulewidth}
\funsigfour{NoisyMax}
{\code{size}$\annotation{:\tyreal_{\pair 0 0}}$, \code{q}$\annotation{:\tylist~\tyreal_{\pair * *}}$}
{\code{max}$\annotation{:\tyreal_{\pair 0 -}}$}
{$\forall$ \code{i}.~ $-1\leq$~\code{$\first{\distance{\text{q}}}$[i]}~$\leq1$ $\land$ $\second{\distance{\code{q}}}\code{[i]} = \first{\distance{\code{q}}}\code{[i]}$}
\algrule

\begin{lstlisting}[frame=none, escapechar=@,basicstyle=\appendixalgsize\ttfamily]
i := 0; bq := 0; max := 0;
while (i < size)
  $\eta$ := $\lapm(2/\priv)$;@\label{line:noisymax_eta}@
  if (q[i] + $\eta$ > bq $\lor$ i = 0)@\label{line:noisymax_branch}@
   max := i;@\label{line:noisymax_true_branch}@@\label{line:noisymax_out}@
   bq := q[i] + $\eta$;@\label{line:noisymax_true_branch_2}@
  i := i + 1;
\end{lstlisting}
\noindent\rule{\linewidth}{0.8pt}
\funsigthree{\small Transformed NoisyMax}
{\code{size,q}\instrument{, \distance{\code{q}}, $\noise$, $\hole$}}
{(\code{max})}
\algrule
\begin{lstlisting}[frame=none, escapechar=@,firstnumber=8,basicstyle=\appendixalgsize\ttfamily]
@\instrument{$\vpriv{\epsilon}$ := 0; idx := 0;}@
i := 0; bq := 0; max := 0;
$\instrument{\first{\distance{\code{bq}}}\text{ := 0;}\quad\second{\distance{\code{bq}}}\text{ := 0;}\quad\first{\distance{\code{max}}} \text{ := 0;}\quad\second{\distance{\code{max}}}\text{ := 0;}}$ /*\quad\second{\distance{\code{max}}}\text{ := 0;}*/
while (i < size)
  @\instrument{$\eta$ := $\noise$[idx]; $\vpriv{\epsilon}$ := ($\ternary{\mathcal{S}}{\vpriv{\epsilon}}{0}$) + $|\alignment| \times \priv$;}@
  @\instrument{\distance{$\eta$} := $\alignment$;}@
  @\instrument{\code{\textbf{if}} ($\select$)\quad$\first{\distance{bq}}$ := $\second{\distance{bq}}$; $\first{\distance{max}}$ := $\second{\distance{max}}$;}@@\label{line:noisymax_shadow_update}@
  if (q[i] + $\eta$ > bq $\lor$ i = 0) 
    @\instrument{\assert{$\text{q[i]}+\distance{\text{q}}\text{[i]}+\eta+\first{\eta} > \text{bq} + \first{\text{bq}} \lor \text{i} = 0$};}@
    @\instrument{\assert{$\first{\distance{\code{max}}}$ = 0};}@
    max := i;
    @\instrument{$\first{\code{max}}$ := 0;}@
    $\instrument{\second{\distance{\text{bq}}}\text{ := bq + }\second{\distance{\text{bq}}}\text{ - (q[i] + }\eta\text{);}}$@\label{line:noisymax_bq_shadow}@
    bq := q[i] + $\eta$;
    @\instrument{$\first{\distance{\text{bq}}}$ := $\first{\distance{\text{q}}}$[i] + $\first{\distance{$\eta$}}$;}@
  $\instrument{\code{\textbf{else}}}$
    @\instrument{\assert{$\lnot(\text{q[i]}+\distance{\text{q}}\text{[i]}+\eta+\first{\eta} > \text{bq} + \first{\text{bq}} \lor \text{i} = 0)$};}@  
  // shadow execution@\label{line:noisymax_shadow_exe_start}@
  $\instrument{\code{\textbf{if}}~(\text{q[i]} + \second{\distance{\text{q}}}\text{[i]} + \eta> \text{bq} + \second{\distance{\text{bq}}}\mathbin{\lor}\text{i = 0})}$@\label{line:noisymax_shadow_condition}@
    $\instrument{\second{\distance{\text{bq}}}\text{ := q[i] + }\second{\distance{\text{q}}}\text{[i]} + \eta - \text{bq}\text{;}}$
    @\instrument{$\second{\distance{\text{max}}}$ := i - max;}@@\label{line:noisymax_shadow_exe_end}@
  i := i + 1;
\end{lstlisting}
\noindent\rule{\linewidth}{2\arrayrulewidth}
\caption{Report Noisy Max and its transformed code, where $\mathcal{S}$ = $\ternary{\code{q[i]}+\eta > \code{bq} \lor \code{i = 0}}{\hole[0]}{\hole[1]}$ and $\alignment$ = $\hole[3] + \hole[4] \times \first{\distance{q}}\code{[i]} + \hole[5] \times \first{\code{bq}}$}
\label{alg:noisymax}
\end{figure}

\section{Extra Case Studies}
In this section we list the pseudo-code of the algorithms we evaluated in the paper for completeness. The incorrect part for the incorrect algorithms is marked with a box.

\subsection{Report Noisy Max}

\paragraph*{Report Noisy Max~\cite{Dwork06diffpriv}}
This is an important building block for developing differentially private algorithms. It generates differentially private synthetic data by finding the identity with the maximum (noisy) score in the database. Here we present this mechanism in a simplified manner: for a series of query answers \code{q}, where each of them can differ at most one in the adjacent underlying database, its goal is to return the index of the maximum query answer in a privacy-preserving way. To achieve differential privacy, the mechanism first adds $\eta = \lapm{2/\epsilon}$ noise to each of the query answer, then returns the index of the maximum noisy query answers $\code{q[i]} + \eta$, instead of the true query answers \code{q[i]}. The pseudo code of this mechanism is shown in Figure~\ref{alg:noisymax}. 

To prove its correctness using randomness alignment technique, we need to align the only random variable $\eta$ in the mechanism (Line~\ref{line:noisymax_eta}). Therefore, a corresponding privacy cost of aligning $\eta$ would be incurred for each iteration of the loop. However, manual proof~\cite{Dwork06diffpriv} suggests that we only need to align the random variable added to the actual maximum query answer. In other words, we need an ability to ``reset'' the privacy cost upon seeing a new current maximum noisy query answer.

\paragraph*{Bad Noisy Max}
We also created an incorrect variant of Report Noisy Max. This variant directly returns the maximum noisy query answer, instead of the \emph{index}. 

More specifically, it can be obtained by changing Line~\ref{line:noisymax_out} in Figure~\ref{alg:noisymax} from \code{max := i} to \code{max := q[i] + $\eta$}. \tool is then able to find a counterexample for this incorrect variant.

\subsection{Variants of Sparse Vector Technique}

\paragraph*{SVT}

We first show a correctly-implemented standard version of SVT~\cite{ninghuisparse}. This standard implementation is less powerful than running example GapSVT, as it outputs $\true$ instead of the gap between noisy query answer and noisy threshold. This can be obtained by changing Line~\ref{line:gapsvt_output} in Figure~\ref{alg:gapsvt} from \code{out := (q[i] + $\eta_2$)::out;} to \code{out := $\true$::out;}.

\paragraph*{SVT with Monotonic Queries}
There exist use cases with SVT where the queries are monotonic. More formally, queries are monotonic if for related queries $q \sim q'$, $\forall i.~q_i \le q_i'$ or $\forall i.~q_i \ge q_i'$. As shown in~\cite{ninghuisparse}. When the queries are monotonic, it suffices to add $\eta_2\code{ := }\lapm{2N/\priv}$ to each queries (Line~\ref{line:gapsvt_eta2} in Figure~\ref{alg:gapsvt}) and the algorithm still satisfies $\priv$-DP.

Thanks to the flexibility of \tool, it only requires one change in the function specification in order to verify this variant: modify the constraint on $\distance{\code{q}}\code{[i]}$ in the precondition. Specifically, the new precondition for SVT with monotonic queries becomes
$\forall$ \code{i}.~ $0\leq$~\code{\distance{q}[i]}~$\leq1$ for the $\forall i.~q_i \le q_i'$ and $\forall$ \code{i}.~ $-1\leq$~\code{\distance{q}[i]}~$\leq0$ for the other case. The final found alignment by \tool is the same as the ones reported in the manual randomness alignment based proofs~\cite{freegap}: 
$$
\eta_1: 0 \quad \eta_2: \begin{cases}
\code{q[i]}+\eta_2\ge\tT \mathbin{?} 1-\code{\distance{q}[i]}\mathbin{:}0, \text{ if $\forall i.~q_i \le q_i'$} \cr
\code{q[i]}+\eta_2\ge\tT \mathbin{?} -\code{\distance{q}[i]}\mathbin{:}0,\quad\  \text{otherwise}
\end{cases}
$$

To the best of our knowledge, no prior verification works have automatically verified this variant.

\begin{figure}[!ht]
\setstretch{0.9}
\raggedright
\small
\noindent\rule{\linewidth}{0.8pt}
\funsigfour{SVT}
{\code{T,N,size}\annotation{:\tyreal_0}, $q$\annotation{:\tylist~\tyreal_*}}
{(\code{out}\annotation{:\tylist~\bool}), \bound{\priv}}
{\alldiffer}
\algrule
\begin{lstlisting}[frame=none,escapechar=@,basicstyle=\appendixalgsize\ttfamily]
$\eta_1$ := $\lapm{(2/\priv)}$
$\tT$ := $T + \eta_1$;
count := 0; i := 0;
while (count < N $\land$ i < size)
  $\eta_2$ := $\lapm{(4N/\priv)}$
  if ($\text{q[i]}+\eta_2\geq \tT$) then
    out := true::out;
    count := count + 1;
  else
    out := false::out;
  i := i + 1;
\end{lstlisting}
\noindent\rule{\linewidth}{0.8pt}
\funsigthree{Transformed SVT}
{\code{T,N,size,q}\instrument{, \code{\distance{q}}, $\noise$, $\hole$}}
{(\code{out})}
\algrule
\begin{lstlisting}[frame=none,escapechar=@,firstnumber=12,basicstyle=\appendixalgsize\ttfamily]
@\instrument{$\vpriv{\priv}$ := 0; idx = 0;}@
@\instrument{$\eta_1$ := $\noise$[idx]; idx := idx + 1;}@
@\instrument{$\vpriv{\epsilon}$ := $\vpriv{\epsilon}$ + $|\alignment_1| \times \priv / 2$; \distance{$\eta_1$} := $\alignment_1$;}@
$\tT$ := $T$ + $\eta_1$;
@\instrument{\distance{$\tT$} := \distance{$\eta_1$};}@
count := 0; i := 0;
while (count < N $\land$ i < size)
  @\instrument{$\eta_2$ := $\noise$[idx]; idx := idx + 1;}@
  @\instrument{$\vpriv{\priv}$ := $\vpriv{\priv}$ + $|\alignment_2| \times \priv / 4N$; \distance{$\eta_2$} := $\alignment_2$;}@
  if (q[i] + $\eta_2$ $\geq$ $\tT$) then
    @\instrument{\assert{q[i] + $\eta_2$ + \distance{q}[i] + \distance{$\eta_2$} $\geq$ $\tT$ + \distance{$\tT$}};}@
    out := true::out;
    count := count + 1;
  else
    @\instrument{\assert{$\lnot$(q[i] + $\eta_2$ + \distance{q}[i] + \distance{$\eta_2$} $\ge$ $\tT$ + \distance{$\tT$})};}@
    out := false::out;
  i := i + 1;
@\instrument{\assert{$\vpriv{\priv} \le \priv$};}@
\end{lstlisting}
\noindent\rule{\linewidth}{0.8pt}
\caption{Standard Sparse Vector Technique and its transformed code, where underlined parts are added by \tool. The transformed
code contains two alignment templates for $\eta_1$ and $\eta_2$:
$\alignment_1 = \hole[0]$ and $\alignment_2 = (\code{q[i] +
$\eta_2$[i] $\ge$ $\tT$})\mathbin{?}(\hole[1] + \hole[2] \times
\distance{$\tT$} + \hole[3] \times
\distance{\code{q}}\code{[i]})\mathbin{:}(\hole[4] + \hole[5] \times \tT +
\hole[6] \times \distance{\code{q}}\code{[i]})$.\label{alg:svt}}
\end{figure}

\paragraph*{NumSVT}

Numerical Sparse Vector (NumSVT)~\cite{diffpbook} is another interesting correct variant of SVT which outputs a numerical answer when the input query is larger than the noisy threshold. It follows the same procedure as Sparse Vector Technique, the difference is that it draws a fresh noise $\eta_3$ in the $\true$ branch, and outputs $q[i] + \eta_3$ instead of $\true$. Note that this is very similar to our running example GapSVT and BadGapSVT, the key difference is that the freshly-drawn random noise hides the information about $\tT$, unlike the BadGapSVT. This variant can be obtained by making the following changes in Figure~\ref{alg:gapsvt}: (1) Line~\ref{line:gapsvt_eta_1} is changed from $\lapm{2 / \epsilon}$ to $\lapm{3 / \epsilon}$; (2) Line~\ref{line:gapsvt_eta2} is changed from $\lapm{4N / \epsilon}$ to $\lapm{6N / \epsilon}$; (3) Line~\ref{line:gapsvt_output} is change from \code{out := (q[i] + $\eta$)::out;} to ``\code{$\eta_3$ := $\lapm{(3N/\priv)}$; out := (q[i] + $\eta_3$)::out;}''. \tool finds the same alignment as shown in~\cite{lightdp} with which CPAChecker is able to verify the algorithm with this generated alignment.

\paragraph*{Adaptive SVT}

As mentioned in Section~\ref{sec:impexp}, we list the pseudo code of Adaptive SVT in Figure~\ref{alg:adaptivesvt}.

\paragraph*{BadSVT1 - 3}

We now study other three incorrect variants of SVT collected from~\cite{ninghuisparse}. All three variants are based on the classic SVT algorithm we have seen (i.e., Line~\ref{line:gapsvt_output} in Figure~\ref{alg:gapsvt} is \code{out := $\true$::out;}).

BadSVT1~\cite{stoddard2014differentially} adds no noise to the query answers and has no bounds on the number of $\true$'s it can output. This variant is obtained by changing Line~\ref{line:gapsvt_while} from \code{\textbf{while}} (\code{count<N}$\land$\code{i<size}) to \code{\textbf{while}} (\code{i<size}) and Line~\ref{line:gapsvt_eta2} from $\lapm{4N/\epsilon}$ to $0$.
Another variant BadSVT2~\cite{chen2015differentially} has no bounds on the number of $\true$'s it can output as well. It keeps outputting $\true$ even if the given privacy budget has been exhausted. Moreover, the noise added to the queries does not scale with parameter \code{N}. Specifically, based on BadSVT1, Line~\ref{line:gapsvt_eta2} is changed to $\lapm{2 / \epsilon}$. BadSVT3~\cite{lee2014top} is an interesting case since it tries to spend its privacy budget in a different allocation strategy between the threshold \code{T} and the query answers \code{q[i]} ($1:3$ instead of $1:1$). However, the noise added to $\eta_2$ does not scale with parameter \code{N}. The $3/4$ privacy budget is allocated to each of the queries where it should be shared among them. To get this variant, based on SVT  algorithm, the noise generation commands (Line~\ref{line:gapsvt_eta_1}  and Line~\ref{line:gapsvt_eta2}) are changed to \code{$\eta_1$ := $\lapm{4/\epsilon}$} and \code{$\eta_2$ := $\lapm{4/(3\times \epsilon)}$ }, respectively.

\begin{figure}[H]
\setstretch{0.9}
\raggedright
\small
\noindent\rule{\linewidth}{0.8pt}
\funsigfour{NumSVT}
{\code{T,N,size}\annotation{:\tyreal_0}, $q$\annotation{:\tylist~\tyreal_*}}
{(\code{out}\annotation{:\tylist~\bool}), \bound{\priv}}
{\alldiffer}
\algrule
\begin{lstlisting}[frame=none,escapechar=@,basicstyle=\appendixalgsize\ttfamily]
$\eta_1$ := $\lapm{(3/\priv)}$
$\tT$ := $T + \eta_1$;
count := 0; i := 0;
while (count < N $\land$ i < size)
  $\eta_2$ := $\lapm{(6N/\priv)}$
  if ($\text{q[i]}+\eta_2\geq \tT$) then
    $\eta_3$ := $\lapm{(3N/\priv)}$;
    out := (q[i] + $\eta_3$)::out;
    count := count + 1;
  else
    out := false::out;
  i := i + 1;
\end{lstlisting}
\noindent\rule{\linewidth}{0.8pt}
\funsigthree{Transformed NumSVT}
{\code{T,N,size,q}\instrument{, \distance{q}, $\noise$, $\hole$}}
{(\code{out})}
\algrule
\begin{lstlisting}[frame=none,escapechar=@,firstnumber=12,basicstyle=\appendixalgsize\ttfamily]
@\instrument{$\vpriv{\priv}$ := 0; idx = 0;}@
@\instrument{$\eta_1$ := $\noise$[idx]; idx := idx + 1;}@
@\instrument{$\vpriv{\epsilon}$ := $\vpriv{\epsilon}$ + $|\alignment_1| \times \priv / 3$; \distance{$\eta_1$} := $\alignment_1$;}@
$\tT$ := $T$ + $\eta_1$;
@\instrument{\distance{$\tT$} := \distance{$\eta_1$};}@
count := 0; i := 0;
while (count < N $\land$ i < size)
  @\instrument{$\eta_2$ := $\noise$[idx]; idx := idx + 1;}@
  @\instrument{$\vpriv{\priv}$ := $\vpriv{\priv}$ + $|\alignment_2| \times \priv / 6N$; \distance{$\eta_2$} := $\alignment_2$;}@
  if (q[i] + $\eta_2$ $\geq$ $\tT$) then
    @\instrument{\assert{q[i] + $\eta_2$ + \distance{q}[i] + \distance{$\eta_2$} $\geq$ $\tT$ + \distance{$\tT$}};}@
    @\instrument{$\eta_3$ := $\noise$[idx]; idx := idx + 1;}@
    @\instrument{$\vpriv{\priv}$ := $\vpriv{\priv}$ + $|\alignment_3| \times \priv / 3N$; \distance{$\eta_3$} := $\alignment_3$;}@
    @\instrument{\assert{\distance{q}[i] + \distance{$\eta_3$} = 0};}@
    out := (q[i] + $\eta_3$)::out;
    count := count + 1;
  else
    @\instrument{\assert{$\lnot$(q[i] + $\eta_2$ + \distance{q}[i] + \distance{$\eta_2$} $\ge$ $\tT$ + \distance{$\tT$})};}@
    out := false::out;
  i := i + 1;
@\instrument{\assert{$\vpriv{\priv} \le \priv$};}@
\end{lstlisting}
\noindent\rule{\linewidth}{0.8pt}
\caption{Numerical Sparse Vector Technique and its transformed code, where underlined parts are added by \tool. The transformed
code contains three alignment templates for $\eta_1$, $\eta_2$ and $\eta_3$ respectively:
$\alignment_1 = \hole[0]$, $\alignment_2 = (\code{q[i] +
$\eta_2$[i] $\ge$ $\tT$})\mathbin{?}(\hole[1] + \hole[2] \times
\distance{$\tT$} + \hole[3] \times
\distance{\code{q}}\code{[i]})\mathbin{:}(\hole[4] + \hole[5] \times \tT +
\hole[6] \times \distance{\code{q}}\code{[i]})$, $\alignment_3 = \hole[7] + \hole[8] \times
\distance{$\tT$} + \hole[9] \times
\distance{\code{q}}\code{[i]})$\label{alg:numsvt}}
\end{figure}

\begin{figure}[H]
\setstretch{0.8}
\raggedright
\small
\noindent\rule{\linewidth}{0.8pt}
\funsigfour{AdaptiveSVT}
{\code{T,N,size}\annotation{:\tyreal_{0}},\code{q}\annotation{:\tylist~\tyreal_{*}}}
{(\code{out}\annotation{:\tylist~\tyreal_{0}}), \bound{\priv}}
{\alldiffer}
\algrule
\begin{lstlisting}[frame=none,escapechar=@]
cost := 0;
$\eta_1$ := $\lapm{(2/\priv)}$;
cost := cost + $\epsilon/2$;
$\tT$ := $T + \eta_1$;
i := 0;
while (cost $\le$ $\epsilon$ - $4N / \epsilon$ $\land$ i < size)
  $\eta_2$ := $\lapm{(8N/\priv)}$;
  if ($\text{q[i]}+\eta_2 - \tT\geq \sigma$) then
    out := (q[i] + $\eta_2$ - $\tT$)::out;
    cost := cost + $\priv / (8N)$;
  else
    $\eta_3$ := $\lapm{(4N/\priv)}$;
    if ($\text{q[i]}+\eta_3 - \tT\geq 0$) then
      out := (q[i] + $\eta_3$ - $\tT$)::out;
      cost := cost + $\priv / (4N)$;
    else
      out := 0::out;
  i := i + 1;
\end{lstlisting}

\noindent\rule{\linewidth}{0.8pt}
\funsigthree{Transformed AdaptiveSVT}
{\code{T,N,size,q}\instrument{, \code{\distance{q}}, $\noise$, $\hole$}}
{(\code{out})}
\algrule
\begin{lstlisting}[frame=none,escapechar=@,firstnumber=12]
@\instrument{$\vpriv{\priv}$ := 0; idx = 0;}@
@\instrument{$\eta_1$ := $\noise$[idx]; idx := idx + 1;}@
@\instrument{$\vpriv{\epsilon}$ := $\vpriv{\epsilon}$ + $|\alignment_1| \times \priv / 2$; \distance{$\eta_1$} := $\alignment_1$;}@
$\tT$ := $T$ + $\eta_1$;
@\instrument{\distance{$\tT$} := \distance{$\eta_1$};}@
count := 0; i := 0;
while (cost $\le$ $\epsilon$ - $4N / \epsilon$ $\land$ i < size)
  @\instrument{$\eta_2$ := $\noise$[idx]; idx := idx + 1;}@
  @\instrument{$\vpriv{\priv}$ := $\vpriv{\priv}$ + $|\alignment_2| \times \priv / 8N$; \distance{$\eta_2$} := $\alignment_2$;}@
  if (q[i] + $\eta_2$ - $\tT$ $\geq$ $\sigma$) then
    @\instrument{\assert{q[i] + $\eta_2$ + \distance{q}[i] + \distance{$\eta_2$} - ($\tT$ + \distance{$\tT$}) $\geq$ $\sigma$};}@
    @\instrument{\assert{\distance{q}[i] + \distance{$\eta_2$} - \distance{$\tT$} == 0};}@
    out := (q[i] + $\eta_2$ - $\tT$)::out;
    cost := cost + $\priv/(8N)$;
  else
    @\instrument{\assert{$\lnot$(q[i] + $\eta_2$ + \distance{q}[i] + \distance{$\eta_2$} - ($\tT$ + \distance{$\tT$}) $\geq$ $\sigma$)};}@
    @\instrument{$\eta_3$ := $\noise$[idx]; idx := idx + 1;}@
    @\instrument{$\vpriv{\priv}$ := $\vpriv{\priv}$ + $|\alignment_3| \times \priv / 4N$; \distance{$\eta_2$} := $\alignment_3$;}@
    if (q[i] + $\eta_3$ - $\tT$ $\geq$ 0)
      @\instrument{\assert{q[i] + $\eta_3$ + \distance{q}[i] + \distance{$\eta_3$} - ($\tT$ + \distance{$\tT$} $\ge$ 0} ;}@
      @\instrument{\assert{\distance{q}[i] + \distance{$\eta_3$} - \distance{$\tT$} == 0};}@
      out := (q[i] + $\eta_3$ - $\tT$)::out;
      cost := cost + $\priv/(4N)$;
    else
      @\instrument{\assert{$\lnot$(q[i] + $\eta_3$ + \distance{q}[i] + \distance{$\eta_3$} - ($\tT$ + \distance{$\tT$}) $\ge$ 0)};}@
      out := false::out;
  i := i + 1;
@\instrument{\assert{$\vpriv{\priv} \le \priv$};}@
\end{lstlisting}
\noindent\rule{\linewidth}{0.8pt}
\caption{Adaptive SVT and its transformed code, where underlined parts are added by \tool. The transformed
code contains three alignment templates for $\eta_1$ and $\eta_2$:
$\alignment_1 = \hole[0]$, $\alignment_2 = \Omega_{Top}\mathbin{?}(\hole[1] + \hole[2] \times
\distance{$\tT$} + \hole[3] \times
\distance{\code{q}}\code{[i]})\mathbin{:}(\hole[4] + \hole[5] \times \tT +
\hole[6] \times \distance{\code{q}}\code{[i]})$ and $\alignment_3 = \Omega_{Middle}\mathbin{?}(\hole[1] + \hole[2] \times
\distance{$\tT$} + \hole[3] \times
\distance{\code{q}}\code{[i]})\mathbin{:}(\hole[4] + \hole[5] \times \tT +
\hole[6] \times \distance{\code{q}}\code{[i]})$, where $\Omega_{*}$ denotes the corresponding branch condition at Line 8 and 13.\label{alg:adaptivesvt}}
\end{figure}

Note that apart from BadSVT1, which does not sample $\eta_2$, the generated templates are identical to the GapSVT since they all have similar typing environments.

Interestingly, since the errors are very similar among them (no bounds on number of outputs / wrong scale of added noise), \tool finds a common counterexample $[0, 0, 0, 0, 0], [1, 1, 1, 1, -1]$ where $T=0$ and $N=1$ within 6 seconds, and this counterexample is further validated by PSI.

\paragraph*{BadGapSVT}
As discussed in Section~\ref{sec:overview}, we list one of our running examples BadGapSVT in Figure~\ref{alg:igapsvt} for completeness.

\begin{figure}[ht]
\setstretch{0.9}
\raggedright
\small
\noindent\rule{\linewidth}{0.8pt}
\funsigfour{BadSVT1}
{\code{T,N,size}\annotation{:\tyreal_0}, $q$\annotation{:\tylist~\tyreal_*}}
{(\code{out}\annotation{:\tylist~\bool}), \bound{\priv}}
{\alldiffer}
\algrule
\begin{lstlisting}[frame=none,escapechar=@,basicstyle=\appendixalgsize\ttfamily]
$\eta_1$ := $\lapm{(2/\priv)}$;
$\tT$ := $T + \eta_1$;
count := 0; i := 0;
while (@\fbox{i < size}@)
  $\eta_2$ := @\fbox{0}@;
  if ($\text{q[i]}+\eta_2\geq \tT$) then
    out := true::out;
    count := count + 1;
  else
    out := false::out;
  i := i + 1;
\end{lstlisting}
\noindent\rule{\linewidth}{0.8pt}
\funsigthree{Transformed BadSVT1}
{\code{T,N,size,q}\instrument{, \code{\distance{q}}, $\noise$, $\hole$}}
{(\code{out})}
\algrule
\begin{lstlisting}[frame=none,escapechar=@,firstnumber=12,basicstyle=\appendixalgsize\ttfamily]
@\instrument{$\vpriv{\priv}$ := 0; idx = 0;}@
@\instrument{$\eta_1$ := $\noise$[idx]; idx := idx + 1;}@
@\instrument{$\vpriv{\epsilon}$ := $\vpriv{\epsilon}$ + $|\alignment_1| \times \priv / 2$; \distance{$\eta_1$} := $\alignment_1$;}@
$\tT$ := $T$ + $\eta_1$;
@\instrument{\distance{$\tT$} := \distance{$\eta_1$};}@
count := 0; i := 0;
while (i < size)
  $\eta_2$ := 0;
  if (q[i] + $\eta_2$ $\geq$ $\tT$) then
    @\instrument{\assert{q[i] + $\eta_2$ + \distance{q}[i] $\geq$ $\tT$ + \distance{$\tT$}};}@
    out := true::out;
    count := count + 1;
  else
    @\instrument{\assert{$\lnot$(q[i] + $\eta_2$ + \distance{q}[i] $\ge$ $\tT$ + \distance{$\tT$})};}@
    out := false::out;
  i := i + 1;
@\instrument{\assert{$\vpriv{\priv} \le \priv$};}@
\end{lstlisting}
\noindent\rule{\linewidth}{0.8pt}
\caption{BadSVT1 and its transformed code, where underlined parts are added by \tool. The transformed
code contains a alignment template for $\eta_1$:
$\alignment_1 = \hole[0]$.\label{alg:isvt1}}
\end{figure}

\begin{figure}[H]
\setstretch{0.9}
\raggedright
\small
\noindent\rule{\linewidth}{0.8pt}
\funsigfour{BadSVT2}
{\code{T,N,size}\annotation{:\tyreal_0}, $q$\annotation{:\tylist~\tyreal_*}}
{(\code{out}\annotation{:\tylist~\bool}), \bound{\priv}}
{\alldiffer}
\algrule
\begin{lstlisting}[frame=none,escapechar=@,basicstyle=\appendixalgsize\ttfamily]
$\eta_1$ := $\lapm{(2/\priv)}$;
$\tT$ := $T + \eta_1$;
count := 0; i := 0;
while (@\fbox{i < size}@)
  $\eta_2$ := @\fbox{$\lapm{(2/\priv)}$}@;
  if ($\text{q[i]}+\eta_2\geq \tT$) then
    out := true::out;
    count := count + 1;
  else
    out := false::out;
  i := i + 1;
\end{lstlisting}
\noindent\rule{\linewidth}{0.8pt}
\funsigthree{Transformed BadSVT2}
{\code{T,N,size,q}\instrument{, \code{\distance{q}}, $\noise$, $\hole$}}
{(\code{out})}
\algrule
\begin{lstlisting}[frame=none,escapechar=@,firstnumber=12,basicstyle=\appendixalgsize\ttfamily]
@\instrument{$\vpriv{\priv}$ := 0; idx = 0;}@
@\instrument{$\eta_1$ := $\noise$[idx]; idx := idx + 1;}@
@\instrument{$\vpriv{\epsilon}$ := $\vpriv{\epsilon}$ + $|\alignment_1| \times \priv / 2$; \distance{$\eta_1$} := $\alignment_1$;}@
$\tT$ := $T$ + $\eta_1$;
@\instrument{\distance{$\tT$} := \distance{$\eta_1$};}@
count := 0; i := 0;
while (i < size)
  @\instrument{$\eta_2$ := $\noise$[idx]; idx := idx + 1;}@
  @\instrument{$\vpriv{\priv}$ := $\vpriv{\priv}$ + $|\alignment_2| \times \priv / 2$; \distance{$\eta_2$} := $\alignment_2$;}@
  if (q[i] + $\eta_2$ $\geq$ $\tT$) then
    @\instrument{\assert{q[i] + $\eta_2$ + \distance{q}[i] + \distance{$\eta_2$} $\geq$ $\tT$ + \distance{$\tT$}};}@
    out := true::out;
    count := count + 1;
  else
    @\instrument{\assert{$\lnot$(q[i] + $\eta_2$ + \distance{q}[i] + \distance{$\eta_2$} $\ge$ $\tT$ + \distance{$\tT$})};}@
    out := false::out;
  i := i + 1;
@\instrument{\assert{$\vpriv{\priv} \le \priv$};}@
\end{lstlisting}
\noindent\rule{\linewidth}{0.8pt}
\caption{BadSVT2 and its transformed code, where underlined parts are added by \tool. The transformed
code contains two alignment templates for $\eta_1$ and $\eta_2$:
$\alignment_1 = \hole[0]$ and $\alignment_2 = (\code{q[i] +
$\eta_2$ $\ge$ $\tT$})\mathbin{?}(\hole[1] + \hole[2] \times
\distance{$\tT$} + \hole[3] \times
\distance{\code{q}}\code{[i]})\mathbin{:}(\hole[4] + \hole[5] \times \tT +
\hole[6] \times \distance{\code{q}}\code{[i]})$.\label{alg:isvt2}}
\end{figure}

\begin{figure}[ht]
\setstretch{0.9}
\raggedright
\small
\noindent\rule{\linewidth}{0.8pt}
\funsigfour{BadSVT3}
{\code{T,N,size}\annotation{:\tyreal_0}, $q$\annotation{:\tylist~\tyreal_*}}
{(\code{out}\annotation{:\tylist~\bool}), \bound{\priv}}
{\alldiffer}
\algrule
\begin{lstlisting}[frame=none,escapechar=@,basicstyle=\appendixalgsize\ttfamily]
$\eta_1$ := @\fbox{$\lapm{(4/\priv)}$}@;
$\tT$ := $T + \eta_1$;
count := 0; i := 0;
while (count < N $\land$ i < @size@)
  $\eta_2$ := @\fbox{$\lapm{(4/3\priv)}$}@;
  if ($\text{q[i]}+\eta_2\geq \tT$) then
    out := true::out;
    count := count + 1;
  else
    out := false::out;
  i := i + 1;
\end{lstlisting}
\noindent\rule{\linewidth}{0.8pt}
\funsigthree{Transformed BadSVT3}
{\code{T,N,size,q}\instrument{, \code{\distance{q}}, $\noise$, $\hole$}}
{(\code{out})}
\algrule
\begin{lstlisting}[frame=none,escapechar=@,firstnumber=12,basicstyle=\appendixalgsize\ttfamily]
@\instrument{$\vpriv{\priv}$ := 0; idx = 0;}@
@\instrument{$\eta_1$ := $\noise$[idx]; idx := idx + 1;}@
@\instrument{$\vpriv{\epsilon}$ := $\vpriv{\epsilon}$ + $|\alignment_1| \times \priv / 4$; \distance{$\eta_1$} := $\alignment_1$;}@
$\tT$ := $T$ + $\eta_1$;
@\instrument{\distance{$\tT$} := \distance{$\eta_1$};}@
count := 0; i := 0;
while (count < N $\land$ i < size)
  @\instrument{$\eta_2$ := $\noise$[idx]; idx := idx + 1;}@
  @\instrument{$\vpriv{\priv}$ := $\vpriv{\priv}$ + $|\alignment_2| \times 3\priv / 4$; \distance{$\eta_2$} := $\alignment_2$;}@
  if (q[i] + $\eta_2$ $\geq$ $\tT$) then
    @\instrument{\assert{q[i] + $\eta_2$ + \distance{q}[i] + \distance{$\eta_2$} $\geq$ $\tT$ + \distance{$\tT$}};}@
    out := true::out;
    count := count + 1;
  else
    @\instrument{\assert{$\lnot$(q[i] + $\eta_2$ + \distance{q}[i] + \distance{$\eta_2$} $\ge$ $\tT$ + \distance{$\tT$})};}@
    out := false::out;
  i := i + 1;
@\instrument{\assert{$\vpriv{\priv} \le \priv$};}@
\end{lstlisting}
\noindent\rule{\linewidth}{0.8pt}
\caption{BadSVT3 and its transformed code, where underlined parts are added by \tool. The transformed
code contains two alignment templates for $\eta_1$ and $\eta_2$:
$\alignment_1 = \hole[0]$ and $\alignment_2 = (\code{q[i] +
$\eta_2$ $\ge$ $\tT$})\mathbin{?}(\hole[1] + \hole[2] \times
\distance{$\tT$} + \hole[3] \times
\distance{\code{q}}\code{[i]})\mathbin{:}(\hole[4] + \hole[5] \times \tT +
\hole[6] \times \distance{\code{q}}\code{[i]})$.\label{alg:isvt3}}
\end{figure}

\begin{figure}[H]
\setstretch{0.9}
\raggedright
\small
\noindent\rule{\linewidth}{0.8pt}
\funsigfour{BadGapSVT}
{\code{size,T,N}\annotation{:\tyreal_0}, \code{q}\annotation{:\tylist~\tyreal_*}}
{(\code{out}\annotation{:\tylist~\tyreal_0}), \bound{\priv}}
{\alldiffer}
\algrule
\begin{lstlisting}[frame=none,escapechar=@,basicstyle=\appendixalgsize\ttfamily]
$\eta_1$ := $\lapm{(2/\priv)}$;
$\tT$ := $T$ + $\eta_1$;
count := 0; i := 0;
while (count < N)
$\eta_2$ := $\lapm{(4N/\priv)}$;
if (q[i] + $\eta_2\geq\tT$) then
  out := @\fbox{(q[i] + $\eta_2$)}@::out;
  count := count + 1;
else
  out := 0::out;
i := i + 1;
\end{lstlisting}
\noindent\rule{\linewidth}{0.8pt}
\funsigthree{\footnotesize Transformed BadGapSVT}
{\code{T,N,size,q}\instrument{, \distance{q}, $\noise$, $\hole$}}
{(\code{out})}
\algrule
\begin{lstlisting}[frame=none,escapechar=@,firstnumber=12,basicstyle=\appendixalgsize\ttfamily]
@\instrument{$\vpriv{\priv}$ := 0; idx = 0;}@
@\instrument{$\eta_1$ := $\noise$[idx]; idx := idx + 1;}@
@\instrument{$\vpriv{\epsilon}$ := $\vpriv{\epsilon}$ + $|\alignment_1| \times \priv / 2$; \distance{$\eta_1$} := $\alignment_1$;}@
$\tT$ := $T$ + $\eta_1$;
@\instrument{\distance{$\tT$} := \distance{$\eta_1$};}@
count := 0; i := 0;
while (count < N $\land$ i < size)
@\instrument{$\eta_2$ := $\noise$[idx]; idx := idx + 1;}@
@\instrument{$\vpriv{\priv}$ := $\vpriv{\priv}$ + $|\alignment_2| \times \priv / 4N$; \distance{$\eta_2$} := $\alignment_2$;}@
if (q[i] + $\eta_2$ $\geq$ $\tT$) then
  @\instrument{\assert{q[i] + $\eta_2$ + \distance{q}[i] + \distance{$\eta_2$} $\geq$ $\tT$ + \distance{$\tT$}};}@
  @\instrument{\assert{\distance{q}[i] + \distance{$\eta_2$} = 0};}@
  out := (q[i] + $\eta_2$)::out;
  count := count + 1;
else
  @\instrument{\assert{$\lnot$(q[i] + $\eta_2$ + \distance{q}[i] + \distance{$\eta_2$} $\ge$ $\tT$ + \distance{$\tT$})};}@
  out := 0::out;
i := i + 1;
@\instrument{\assert{$\vpriv{\priv} \le \priv$};}@
\end{lstlisting}
\noindent\rule{\linewidth}{0.8pt}
\caption{BadGapSVT and its transformed code. The transformed
code contains two alignment templates for $\eta_1$ and $\eta_2$:
$\alignment_1 = \hole[0]$ and $\alignment_2 = (\code{q[i] +
$\eta_2$[i] $\ge$ $\tT$})\mathbin{?}(\hole[1] + \hole[2] \times
\distance{$\tT$} + \hole[3] \times
\distance{\code{q}}\code{[i]})\mathbin{:}(\hole[4] + \hole[5] \times \tT +
\hole[6] \times \distance{\code{q}}\code{[i]})$. Note that the random variables
and $\hole$ are inserted as part of the function input.\label{alg:igapsvt}}
\end{figure}

\subsection{Partial Sum}

Next, we study a simple algorithm PartialSum (Figure~\ref{alg:partialsum}) which outputs the sum of queries in a privacy-preserving manner: it directly computes sum of all queries and adds a $\lapm{1/\priv}$ to the final output \code{sum}. Note that similar to SmartSum, it has the same adjacency requirement (only one query can differ by at most one). The alignment is easily found for $\eta$ by \tool which is to ``cancel out'' the distance of \code{sum} variable (i.e., \code{-\distance{sum}}). With the alignment CPAChecker verifies this algorithm.

An incorrect variant for PartialSum called BadPartialSum is created where Line~\ref{line:partialsum_sample} is changed from $1/\priv$ to $1 / (2 \times \priv)$, therefore making it fail to satisfy $\priv$-differential privacy (though it actually satisfies $2\priv$-differential privacy). A counterexample $[0,0,0,0,0], [0,0,0,0,1]$ is found by \tool and further validated by PSI.

\begin{figure}[H]
\raggedright
\small
\setstretch{0.9}
\noindent\rule{\linewidth}{2\arrayrulewidth}
\funsigfour{PartialSum}
{\code{size}\annotation{:\tyreal_0}, \code{q}\annotation{:{\tylist~\tyreal_*}}}
{(\code{out}\annotation{:\tyreal_0}), \bound{\priv}}
{\onediffer}
\algrule
\begin{lstlisting}[frame=none,escapechar=@,basicstyle=\appendixalgsize\ttfamily]
sum := 0; i := 0;
while (i < size)
  sum := sum + q[i];
  i := i + 1;
$\eta$ = $\lapm{(1/\priv)}$;@\label{line:partialsum_sample}@
out := sum + $\eta$;
\end{lstlisting}
\noindent\rule{\linewidth}{0.8pt}
\funsigthree{Transformed PartialSum}
{\code{size,q}\instrument{,\code{\distance{q}}, $\noise$, $\hole$}}
{(\code{out})}
\algrule
\begin{lstlisting}[frame=none,escapechar=@,firstnumber=7,basicstyle=\appendixalgsize\ttfamily]
@\instrument{$\vpriv{\priv}$ := 0; \distance{sum} := 0;}@
sum := 0; i := 0;
while (i < size)
  sum := sum + q[i];
  @\instrument{\distance{sum} := \distance{sum} + \distance{q}[i];}@
  i := i + 1;
@\instrument{$\vpriv{\priv}$ := $\vpriv{\priv}$ + |$\alignment$| $\times$ $\priv$; \distance{$\eta$} := $\boldsymbol{\hole}$;}@
@\instrument{\assert{\distance{sum} + \distance{$\eta$}};}@
out := sum + $\eta$;
@\instrument{\assert{$\vpriv{\priv} \le \priv$};}@
\end{lstlisting}
\noindent\rule{\linewidth}{2\arrayrulewidth}
\caption{PartialSum and its transformation using \tool, where $\alignment$ = $\hole$[0] + $\hole$[1] $\times$ \code{\distance{\text{sum}}} + $\hole$[2] $\times$ \code{\distance{q}[i]}.}
\label{alg:partialsum}
\end{figure}

\subsection{SmartSum and BadSmartSum}
SmartSum~\cite{chan10continual} continually releases
aggregated statistics with privacy protections. For a finite sequence of queries $q[0], q[1], \cdots, q[T]$, where $T$ is the length of $q$, the goal of SmartSum is to release the prefix sum: $q[0], q[0] + q[1], \cdots, \sum_{i=0}^T q[i]$ in a private way. To achieve differential privacy, SmartSum first divides the sequence into non-overlapping blocks $B_0, \cdots, B_l$ with size $M$, then maintains the noisy version of each query and noisy version of the block sum, both by directly adding $\lapm{1/\priv}$ noise. Then to compute the $k^{\text{th}}$ component of the prefix sum sequence $\sum_{i=0}^k q[i]$, it only has to add up the noisy block sum that covers before $k$, plus the remaining $(k+1)\ \code{mod}\ M$ noisy queries. The pseudo code is shown in Figure~\ref{alg:smartsum}. The \textbf{if} branch is responsible for dividing the queries and summing up the block sums (stored in \code{sum} variable), where \textbf{else} branch adds the remaining noisy queries.

Notably, SmartSum satisfies 2$\priv$-differential privacy instead of
$\priv$-differential privacy. Moreover, the adjacency requirement of the inputs
is that only one of the queries can differ by at most one. These two
requirements are specified in the function signature ($\bound{2\priv}$ and
\textbf{precondition}).

An incorrect variant of SmartSum, called BadSmartSum, is obtained by changing
Line~\ref{line:smartsum_eta_1} to \code{$\eta_1\!:=\!0$} in
Figure~\ref{alg:smartsum}. It directly
releases \code{sum + q[i]} without adding any noise (since $\eta_1 = 0$), where
\code{sum} stores the accurate, non-noisy sum of queries (at
Line~\ref{line:smartsum_sumcal}), hence breaking differential privacy. Interestingly, the violation only happens in a rare branch \code{\textbf{if} ((i + 1) mod M = 0)}, where the accurate sum is added to the output list \code{out}. In other words, \code{out} contains mostly private data with only a few exceptions. This rare event makes it challenging for sampling-based tools to find the violation.

\begin{figure}[H]
\raggedright
\setstretch{0.9}
\small
\noindent\rule{\linewidth}{2\arrayrulewidth}
\funsigfour{SmartSum}
{\code{M,T,size}\annotation{:\tyreal_0}, \code{q}\annotation{:\tylist~\tyreal_*}}
{(\code{out}\annotation{:\tylist~\tyreal_0}), \bound{$2\priv$}}
{\onediffer}
\algrule
\begin{lstlisting}[frame=none, escapechar=@]
next := 0; i := 0; sum := 0;
while (i < size $\land$ i $\leq$ T)
  if ((i + 1) mod M = 0) then @\label{line:smartsum_if}@
    $\eta_1$ := $\lapm{(1/\priv)}$;@\label{line:smartsum_eta_1}@
    next := sum + q[i] + $\eta_1$;
    sum := 0;
    out := next::out; 
  else @\label{line:smartsum_else}@
    $\eta_2$ := $\lapm{(1/\priv)}$;
    next:= next + q[i] + $\eta_2$;
    sum := sum + q[i];@\label{line:smartsum_sumcal}@
    out := next::out;
  i := i + 1;
\end{lstlisting}
\noindent\rule{\linewidth}{0.8pt}
\funsigthree{Transformed SmartSum}
{\code{M,T,size,q}\instrument{, \code{\distance{q}}, $\noise$, $\hole$}}
{(\code{out})}
\algrule
\begin{lstlisting}[frame=none, escapechar=@,firstnumber=14]
@\instrument{$\vpriv{\priv}$ := 0; idx := 0;}@
next := 0; i := 0; sum := 0;
@\instrument{\distance{sum} := 0; \distance{next} := 0;}@
while (i < size $\land$ i $\leq$ T)
  if ((i + 1) mod M = 0) then
    @\instrument{$\eta_1$ := $\noise$[idx]; idx := idx + 1;}@
    @\instrument{$\vpriv{\priv}$ := $\vpriv{\priv}$ + |$\alignment_1$| $\times$ $\priv$; \distance{$\eta_1$} := $\alignment_1$;}@
    next := sum + q[i] + $\eta_1$;
    @\instrument{\distance{next} := \distance{sum} + \distance{q}[i] + \distance{$\eta_1$};}@
    sum := 0;
    @\instrument{\distance{sum} := 0;}@
    @\instrument{\assert{\distance{next} = 0};}\label{line:smartsum_true_assert}@
    out := next::out;
  else
    @\instrument{$\eta_2$ := $\noise$[idx]; idx := idx + 1;}@
    @\instrument{$\vpriv{\priv}$ := $\vpriv{\priv}$ + |$\alignment_2$| $\times$ $\priv$; \distance{$\eta_2$} := $\alignment_2$;}@
    next := next + q[i] + $\eta_2$;
    @\instrument{\distance{next} := \distance{next} + \distance{q}[i] + \distance{$\eta_2$};}@
    sum := sum + q[i];
    @\instrument{\distance{sum} := \distance{sum} + \distance{q}[i];}@
    @\instrument{\assert{\distance{next} = 0};}@
    out := next::out;
  i := i + 1;
@\instrument{\assert{$\vpriv{\priv} \le 2\priv$};}@
\end{lstlisting}
\noindent\rule{\linewidth}{2\arrayrulewidth}
\caption{SmartSum and its transformed code. Underlined parts are added by \tool. $\alignment_1$ = $\hole[0]$ + $\hole[1]$ $\times$ \code{\distance{sum}} + $\hole[2]$ $\times$ \code{\distance{q}[i]} + $\hole[3]$ $\times$ \code{\distance{next}} and $\alignment_2$ = $\hole[4]$ + $\hole[5]$ $\times$ \code{\distance{sum}} + $\hole[6]$ $\times$ \code{\distance{q}[i]} + $\hole[7]$ $\times$ \code{\distance{next}}.}
\label{alg:smartsum}
\end{figure}

\balance
\fi
\end{document}